\documentclass[%
 reprint,
 amsmath,amssymb,
 aps,
]{revtex4-2}

\usepackage{xcolor}
\usepackage{graphicx}% Include figure files
\graphicspath{{./Figs/}}
\usepackage{dcolumn}% Align table columns on decimal point
\usepackage{bm}% bold math
\begin{document}
\preprint{APS/123-QED}

\title{Dripping-onto-droplet rheometry of sodium alginate solutions}% Force line breaks with \\

\author{Nada Nazzal}
%\thanks{nada.nazzal@minesparis.psl.eu}
\affiliation{Mines Paris, PSL University, Centre for material forming (CEMEF), UMR CNRS 7635, rue  Claude Daunesse, 06904 Sophia-Antipolis, France}

\author{Martin Drah\'e}
%\thanks{martin.drahe@minesparis.psl.eu}
\affiliation{Mines Paris, PSL University, Centre for material forming (CEMEF), UMR CNRS 7635, rue  Claude Daunesse, 06904 Sophia-Antipolis, France}

\author{Ricardo El Khoury}
%\thanks{ricardo.elkhoury@minesparis.psl.eu}
\affiliation{Mines Paris, PSL University, Centre for material forming (CEMEF), UMR CNRS 7635, rue  Claude Daunesse, 06904 Sophia-Antipolis, France}

\author{Irineu Lopes Palhares Junior}
%\thanks{irineu.palhares@unesp.br}
\affiliation{Departamento de Matem\'atica e Computa\c c\~ao, Faculdade de Ci\^encias e Tecnologia, Universidade Estadual Paulista `Julio de Mesquita Filho', Presidente Prudente, Brazil}

\author{Cassio Oishi}
%\thanks{cassio.oishi@unesp.br}
\affiliation{Departamento de Matem\'atica e Computa\c c\~ao, Faculdade de Ci\^encias e Tecnologia, Universidade Estadual Paulista `Julio de Mesquita Filho', Presidente Prudente, Brazil}

\author{Roney Leon Thompson}
%\thanks{rthompson@mecanica.coppe.ufrj.br}
\affiliation{Department of Mechanical Engineering, COPPE, Universidade Federal do Rio de Janeiro, Centro de Tecnologia, Ilha do Fund\~ao, Rio de Janeiro, RJ 24210-240, Brazil}

\author{Loren J{\o}rgensen}
%\thanks{loren.jorgensen@espci.fr}
\affiliation{Sciences et Ing\'enierie de la Mati\`ere Molle (SIMM), ESPCI Paris, Universit\'e PSL, UMR CNRS 7615, Sorbonne Universit\'e, Paris, France}

\author{C\'ecile Monteux}
%\thanks{cecile.monteux@espci.fr}
\affiliation{Sciences et Ing\'enierie de la Mati\`ere Molle (SIMM), ESPCI Paris, Universit\'e PSL, UMR CNRS 7615, Sorbonne Universit\'e, Paris, France}

\author{Edith Peuvrel-Disdier}
%\thanks{edith.peuvrel-disdier@minesparis.psl.eu}
\affiliation{Mines Paris, PSL University, Centre for material forming (CEMEF), UMR CNRS 7635, rue  Claude Daunesse, 06904 Sophia-Antipolis, France}

\author{Anselmo Pereira}
\thanks{anselmo.soeiro\_pereira@minesparis.psl.eu (lead researcher \& corresponding author)}
\affiliation{Mines Paris, PSL University, Centre for material forming (CEMEF), UMR CNRS 7635, rue  Claude Daunesse, 06904 Sophia-Antipolis, France}%

\date{\today}% It is always \today, today,

%%%%%%%%%%%%%%%%%%%%%%%%%%%%%%%%%%%%%%%%%%%%%%%%%%%%%%%%%%%%%%%%%%%%%%%%%%%%%%
%%%%%%%%%%%%%%%%%%%%%%%%%%%%%%%%%%%%%%%%%%%%%%%%%%%%%%%%%%%%%%%%%%%%%%%%%%%%%%
\begin{abstract}
In this experimental and theoretical study, we assess the extensional relaxation time of sodium alginate solutions by using dripping-onto-droplet capillary breakup rheometry (DoD), e.g., the capillary thinning and breakup of viscoelastic filaments formed following the coalescence of a millimetric-nozzle-generated pendant drop with a lower droplet cap of the same fluid contained in a millimetric pool in ambient air. Hence, we extend the analyses conducted by El Khoury et al. (2026) from Newtonian to viscoelastic fluids. Our approach relies on experiments recorded with a high-speed camera using sodium alginate in deionised water, with alginate concentrations ranging from 0.1\% to 9\% by weight. The results are depicted by considering the dynamics of fluid filament thinning, stress balances, and scaling laws. Extensional relaxation times are resolved from the filament diameter evolution. Three flow regimes are highlighted: capillary-inertial, capillary-elastic, and mixed capillary-inertio-elastic. The findings are summarised in a two-dimensional diagram that correlates the filament breakup time with different flow regimes using the important dimensionless parameter of the problem, e.g., the intrinsic Deborah number (which relates the extensional relaxation time to the characteristic capillary-inertial time). This diagram can be used to quantify both the solution's extensional relaxation time and the liquid/air surface tension solely from filament breakup times.          
\vspace*{0.5cm}

 
\end{abstract}
%%%%%%%%%%%%%%%%%%%%%%%%%%%%%%%%%%%%%%%%%%%%%%%%%%%%%%%%%%%%%%%%%%%%%%%%%%%%%%
%%%%%%%%%%%%%%%%%%%%%%%%%%%%%%%%%%%%%%%%%%%%%%%%%%%%%%%%%%%%%%%%%%%%%%%%%%%%%%

\maketitle

%%%%%%%%%%%%%%%%%%%%%%%%%%%%%%%%%%%%%%%%%%%%%%%%%%%%%%%%%%%%%%%%%%%%%%%%%%%%%%
%%%%%%%%%%%%%%%%%%%%%%%%%%%%%%%%%%%%%%%%%%%%%%%%%%%%%%%%%%%%%%%%%%%%%%%%%%%%%%
\section{Introduction} \label{INTRO}
%%%%%%%%%%%%%%%%%%%%%%%%%%%%%%%%%%%%%%%%%%%%%%%%%%%%%%%%%%%%%%%%%%%%%%%%%%%%%%
%%%%%%%%%%%%%%%%%%%%%%%%%%%%%%%%%%%%%%%%%%%%%%%%%%%%%%%%%%%%%%%%%%%%%%%%%%%%%%
Sodium alginate is a naturally occurring polysaccharide extracted from the cell walls of brown algae \citep{Lee_2012, Sun_2013, Vicini_2015, Castellano_2019}. It is a linear biopolymer composed of $\beta$-D-mannuronic acid (M-blocks) and $\alpha$-L-guluronic acid (G-blocks) residues connected by a $\beta$-(1-4) glycosidic bond, and presents negative charges along its backbone, which typically confer a polyelectrolyte nature to its polymer chains. The latter are often nearly rigid, assuming a rodlike conformation in salt-free solvent due to the strong electrostatic repulsions between the mentioned charges. Nevertheless, these electrostatic forces can be mitigated by reducing the chain length. Consequently, low-molecular-weight sodium alginate in salt-free solvent may behave like a neutral polymer in $\theta$ solvent, i.e., the polymer coils act like ideal chains, assuming a random coil conformation \citep[]{Dodero_2019, Dodero_2020}. 

In the presence of divalent ions, sodium alginate solutions undergo gelation (via ionic crosslinking), resulting in densely interconnected hydrogels \citep{Godefroid_2025}. Such materials have been widely applied in 3D printing, tissue engineering, drug delivery systems, and wound dressings, owing to their tunable mechanical properties and structural similarities with the extracellular matrix \citep{Murphy_2014, Bochenek_2018, Yuk_2022, Ji_2022}. For all the mentioned applications, a weakly viscoelastic liquid becomes a viscoelastic solid as the flow develops and divalent ions diffuse within the alginate solution. To finely control them, it is thus crucial to describe the mechanical behaviour of the used fluid, as well as its time evolution. Logically, the first task in this endeavour is to rheologically characterise the initial alginate solution, which can be quite challenging when using standard shear-based rheometry due to the relatively low polymer stretching induced by viscometric shear flows. Note that the very few works in the literature devoted to the rheological characterisation of alginate solutions mainly focus on the viscous aspect of the material \citep[][]{Dodero_2019, Dodero_2020}. In this connection, uniaxial extensional rheometry can represent an interesting technique to access weak viscoelastic properties, since polymer chains tend to be highly stretched by uniaxial extensional flows \citep{Liang_1994, Entov_1997, McKinley_2000, Tuladhar_2008, Keshavarz_2015, Zinelis_2024}. 

In the present work, we characterise alginate solutions using dripping-onto-droplet rheometry \citep[DoD;][]{Khoury_2026}, i.e., the uniaxial extensional capillary thinning and breakup of viscoelastic filaments formed following the coalescence of a millimetric-nozzle-generated pendant drop with a lower droplet cap of the same fluid contained in a millimetric pool in ambient air (see figure \ref{fig-1}). This rheometric technique relies on capillary stresses to favour the growth of Plateau-Rayleigh instabilities \citep{Plateau_1873, Rayleigh_1878, Rayleigh_1880, Rayleigh_1892}, leading to thinning, destabilisation, and breakup of the fluid filament. Different from other liquid-bridge-based techniques such as Capillary Breakup Extensional Rheometer \citep[CaBER;][]{Gaudet_1996, Anna_2001, Rodd_2005, Clasen_2006b, Valette_2019, Joseph_2025}, Dripping-onto-Substrate \citep[DoS;][]{Dinic_2015, Dinic_2017, Rosello_2019}, and Acoustically-Driven Microfluidic Extensional Rheometer \citep[ADMiER;][]{Bhattacharjee_2011, McDonnell_2015}, DoD is a wetting-independent flow configuration in which capillary thinning naturally occurs when the upper drop gently touches the lower one (no fast pre-stretching required). The experiments are recorded with a high-speed camera and using solutions with alginate concentrations ranging from 0.1\% to 9\% by weight in deionised water. Part of their thinning process results from a competition between capillary and elastic stresses leading to a filament diameter $d(t)$ that decays exponentially in time $t$, according to the following equation derived by Entov and Hinch \citep{Entov_1997} for an Oldroyd-B/Hookean dumbbell \citep{Oldroyd_1950, Bird_1987, Hinch_2021}:           
\begin{equation}
\frac{d(t)}{d_0} = A \left[\frac{\left(\eta_0 - \eta_s \right) d_0}{\sigma \lambda_e} \right]^{1/3} \exp{\left( -\frac{t}{3\lambda_e} \right)}  \, ,
\label{eq:intro-1}
\end{equation} 
where $d_0$ is the initial diameter of the pendant drop, $\eta_0$ is the solution's zero-shear viscosity, $\eta_s$ is the solvent viscosity, $\eta_0 - \eta_s$ is the polymeric viscosity $\eta_p$, $\sigma$ is the surface tension, $\lambda_e$ is the extensional relaxation time, and $A=(1/4)^{1/3}$ is a constant \citep{Bazilevskii_1997, Clasen_2006a, Pereira_2013}. Hence, we use equation \ref{eq:intro-1} to access the material's extensional relaxation time. More importantly, we also show that $\lambda_e$ can be directly extracted from the filament breakup time $t_b$ (necessary time to break the filament apart by capillarity after the drops' coalescence) through a master curve connecting $t_b$ to the intrinsic Deborah number defined as $\mathrm{De} = \lambda_e \sqrt{\sigma/(\rho d_0^3)}$ ($\rho$ being the solution's density). 

The paper is organised as follows: a detailed description of the physical formulation and the experimental method is presented in section \ref{PFEMDN}; the important dimensionless parameters of the problem are equally highlighted; results are discussed in section \ref{RD}; finally, conclusions and perspectives are drawn in the closing section.

%%%%%%%%%%%%%%%%%%%%%%%%%%%%%%%%%%%%%%%%%%%%%%%%%%%%%%%%%%%%%%%%%%%%%%%%%%%%%%
%%%%%%%%%%%%%%%%%%%%%%%%%%%%%%%%%%%%%%%%%%%%%%%%%%%%%%%%%%%%%%%%%%%%%%%%%%%%%%
\section{Physical Formulation, Experimental Method, and Dimensionless Numbers} \label{PFEMDN}
%%%%%%%%%%%%%%%%%%%%%%%%%%%%%%%%%%%%%%%%%%%%%%%%%%%%%%%%%%%%%%%%%%%%%%%%%%%%%%
%%%%%%%%%%%%%%%%%%%%%%%%%%%%%%%%%%%%%%%%%%%%%%%%%%%%%%%%%%%%%%%%%%%%%%%%%%%%%%

\begin{figure*}[htp]
\centering    
\includegraphics[width=1\linewidth]{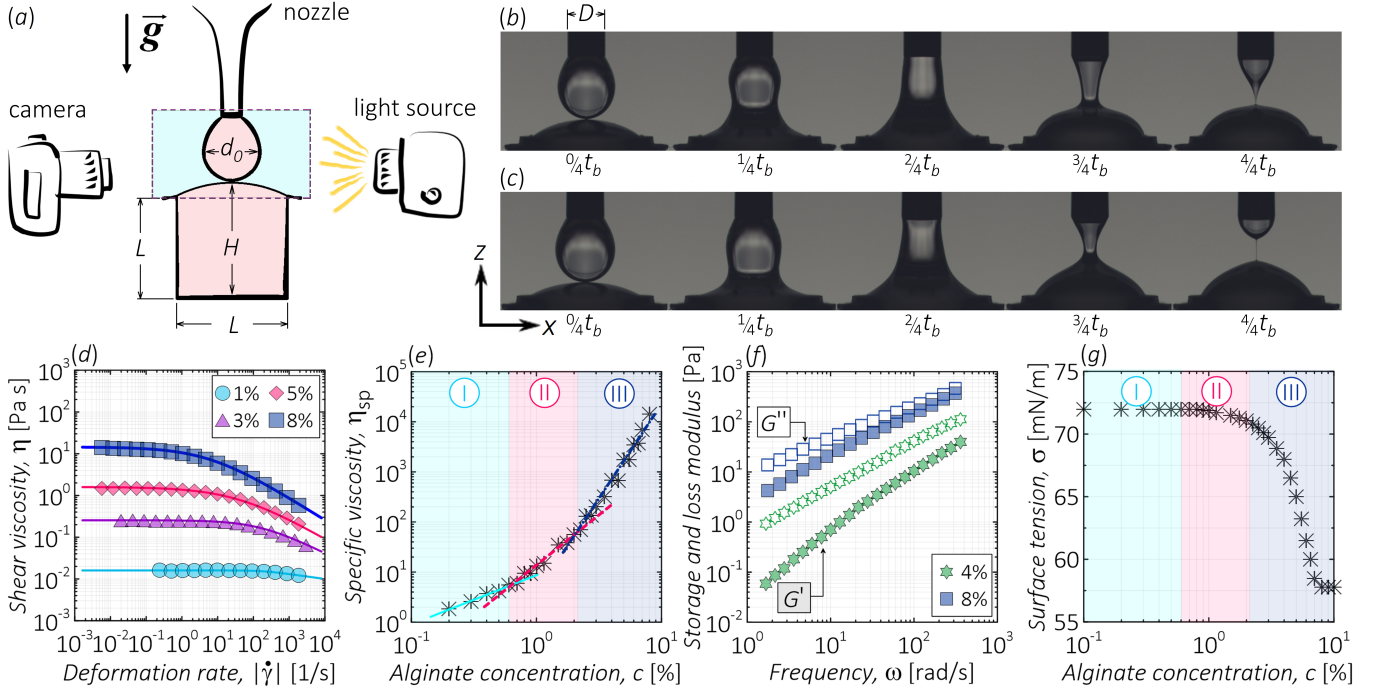}
\vspace{-0.7cm}    
\caption{{\color{darkgray}{Schematic illustration of the experimental set-up for the investigation of the capillary thinning and breakup of fluid filaments formed following the coalescence of a millimetric-nozzle-generated pendant drop (of density $\rho$, viscosity $\eta$, surface tension $\sigma$ and diameter $d_0$) with a lower droplet cap contained in a 3D-printed-polylactic-acid cylindrical pool of millimetric diameter $L$ and height $L$ in ambient air under gravity, i.e., \textit{dripping-onto-droplet capillary breakup} (DoD). The blue rectangle, delimited by the magenta-dashed lines, highlights the free-surface flow region. (\textit{b}-\textit{c}) Typical snapshots captured by the high-speed camera illustrating the capillary-thinning process from the droplets' coalescence until the filament breakup at instant $t_b$. (\textit{d}) Flow sweep tests: shear viscosity $\eta$ plotted as a function of the deformation rate $|\boldsymbol{\dot{\gamma}}|$. Each symbol represents a specific alginate concentration, while the lines are the corresponding Carreau-Yasuda fits. (\textit{e}) Specific shear viscosity $\eta_{sp}$ against the alginate concentration $c$. The data are categorised into three concentration regimes: dilute ($c <$ 0.65\%; cyan zone; I); semi-dilute unentangled (0.65\% $\leq c \leq$ 2.2\%; pink zone; II); and semi-dilute entangled (navy blue zone; III). The points are fitted by a power-law function $\eta_{sp} \propto c^{\alpha}$ whose exponent varies with the concentration regime: $\alpha$ depends on the concentration regime: $\alpha =$ 1 (solid line), 2 (dashed line), and 3.75 (dash-dotted line) in the diluted, semi-diluted unentangled, and semi-diluted entangled concentration regimes, respectively. (\textit{f}) Frequency sweep tests: the storage modulus $G^{\prime}$ (solid symbols) and the loss modulus $G^{\prime \prime}$ (open symbols) are plotted against the oscillation frequency $\omega$ within the linear viscoelastic limit for two alginate concentrations: 4\% (green stars) and 8\% (blue rectangles). (\textit{g}) Surface tension $\sigma$ against the alginate concentration $c$ (dataset obtained by using the pendant drop method).}}}
\vspace{-0.2cm} 
\label{fig-1}
\end{figure*}

As mentioned in the previous section, this study presents a combined experimental and theoretical analysis of the capillary-driven filament thinning process followed by the coalescence between an upper viscoelastic droplet of density $\rho$, shear-thinning viscosity $\eta$, and diameter $d_0$ and a lower droplet of the same fluid, as depicted in sub-figures \ref{fig-1}(\textit{a})-\ref{fig-1}(\textit{c}). The former is generated by a nozzle of millimetric diameter $D$, while the latter is formed when filling a 3D-printed-polylactic-acid cylindrical pool of millimetric diameter and height $L \times L$ (see sub-figure \ref{fig-1}\textit{a}). The droplets' coalescence leads to the formation of a fluid filament whose diameter $d(t)$ progressively decreases under capillary forces until its rupture at instant $t_b$ due to the growth of Plateau-Rayleigh instabilities \citep{Eggers_1993, Eggers_2008, Pita_2012, Deblais_2018, Kooij_2025, Khoury_2026}. The surrounding air is characterised by a density $\rho_{air}$ and viscosity $\eta_{air}$. Both liquid and gas phases are at fixed ambient temperature $T$, and the surface tension between them $\sigma$ is constant. 

As schematised in sub-figure \ref{fig-1}(\textit{a}), the capillary-thinning process is recorded by an Optronis Cyclone 2-2000 high-speed camera operated at $\mathcal{O} \left( 10^{4} \right)$ frames/s and equipped with a Sigma 105mm F2.8 DG OS HSM macro lens. A LED backlight system provides the necessary lighting level. Typical experimental snapshot sequences illustrating the fluid filament breakup are given in sub-figures \ref{fig-1}(\textit{b}) and \ref{fig-1}(\textit{c}).

Sodium-alginate-in-deionised-water solutions are used at twenty-eight alginate concentrations $c$: 0.1\%, 0.2\%, 0.3\%, 0.4\%, 0.5\%, 0.6\%, 0.7\%, 0.8\%, 0.9\%, 1\%, 1.15\%, 1.5\%, 1.75\%, 2\%, 2.25\%, 2.5\%, 2.75\%, 3\%, 3.5\%, 4\%, 4.5\%, 5\%, 5.5\%, 6\%, 6.5\%, 7\%, 8\% and 9\% by weight. The alginate is purchased from Sigma-Aldrich with a molar mass of approximately 100kDa and an M-block/G-block ratio of about 1.56 \citep{Dodero_2019, Dodero_2020}. The solutions are rheologically characterised at 25$^\circ$C using a MCR 302 rheometer by Anton Paar equipped with a cone-plate geometry with 50mm diameter and 2$^\circ$ angle. Steady-shear flow sweep and frequency sweep tests are performed after an initial 3-minute pre-shear step at 1s$^{-1}$. As indicated in sub-figure \ref{fig-1}(\textit{d}), where the shear viscosity $\eta$ is plotted as a function of the deformation rate $|\boldsymbol{\dot{\gamma}}|$, the solutions exhibit a shear-thinning behaviour that becomes more pronounced with an increasing alginate concentration. This tendency results from chain overlaps and entanglements, which make the solution's deformation more energetically costly. Nevertheless, as the deformation rate increases, the polymer chains tend to disentangle and stretch, leading to a decrease in shear viscosity \citep{Kulicke_1982, Kulicke_1984, Xu_2014, Torres_2014, Costanzo_2016, Bertasa_2020}. The latter can be fitted by the Carreau-Yasuda equation \citep{Carreau_1972, Yasuda_1979, Bird_1987}
\begin{equation}
\frac{\eta - \eta_{\infty}}{\eta_0 - \eta_{\infty}} = \left[  1 + \left( \lambda_{CY} |\boldsymbol{\dot{\gamma}}| \right)^a \right]^{\frac{n-1}{a}}  \, ,
\label{eq:etacy}
\end{equation} 
in which $\eta_0$ is the zero-shear viscosity, $\eta_{\infty}$ is the infinite-shear viscosity, $\lambda_{CY}$ is a characteristic time, $n$ is the power-law index, and $a$ is a constant associated with the transition between the upper viscosity plateau ($\eta_0$) and the power-law part of the curve. The zero-shear viscosity and the infinite-shear viscosity (e.g., the solvent viscosity) are used to calculate the specific shear viscosity $\eta_{sp}$
\begin{equation}
\eta_{sp} = \frac{\eta_0 - \eta_{\infty}}{\eta_{\infty}}  \, ,
\label{eq:etasp}
\end{equation} 
which, in turn, is plotted as a function of the alginate concentration in sub-figure \ref{fig-1}(\textit{e}). The data are categorised into three concentration regimes: dilute ($c <$ 0.65\%; cyan zone), in which single polymer chains do not interact with each other; semi-dilute unentangled (0.65\% $\leq c \leq$ 2.2\%; pink zone), in which the chains overlap but entaglements do not occur; and semi-dilute entangled (navy blue zone), in which the chains are no longer isolated and start to interpenetrate each other creating a network. In each of them, the data are fitted by a power-law function, 
\begin{equation}
\eta_{sp} \propto c^{\alpha}  \, ,
\label{eq:etasp}
\end{equation} 
as pointed out by the lines. As observed, the \textit{scaling factor} $\alpha$ depends on the concentration regime: $\alpha =$1 (solid line), 2 (dashed line), and 3.75 (dash-dotted line) in the diluted, semi-diluted unentangled, and semi-diluted entangled concentration regimes, respectively. Curiously, these values are in agreement with those predicted for neutral polymers in $\theta$ solvent \citep{Dobrynin_1995, Kulicke_2004, Colby_2010}. In other words, the polymer coils act like ideal chains, assuming a random coil conformation  \citep[see also][]{Dodero_2019, Dodero_2020}. The transitions between the concentration regimes are characterised by the following critical concentrations given by the intersection between consecutive lines: $c^{\ast} =$ 0.65\%, at which polymer chains start to overlap without entanglement (e.g., the \textit{overlap concentration}); and $c_{en} =$ 2.2\%, above which the polymer chains entangle with each other. Interestingly, as indicated by sub-figure \ref{fig-1}(\textit{f}), no crossover between the storage modulus $G^{\prime}$ and the loss modulus $G^{\prime \prime}$ plotted against the oscillation frequency $\omega$ within the linear viscoelastic limit \citep{Bird_1987} occurs for $c \leq 8\%$. Hence, relaxation times, which are typically obtained from the intersection of $G^{\prime}(\omega)$ and $G^{\prime \prime}(\omega)$, cannot be provided by such a rheological test. This constraint will be overcome using DoD in the following section. Lastly, pendant drop results giving access to the surface tension are illustrated in sub-figure \ref{fig-1}(\textit{g}), where $\sigma$ is displayed as a function of $c$ \citep{Stauffer_1965, Adamson_1997, Ebnesajjad_2011, Berry_2015}. As the alginate concentration is maximised, the amount of polymers at the liquid/air interface increases, consequently reducing the surface tension until saturation is reached at $c \approx 7.7\%$.  

The estimated Carreau-Yasuda parameters, as well as density and surface tension values, are summarised in Table \ref{tb:table-1} for each considered alginate solution at 25$^\circ$C.   
%\vspace{0.5cm}
\begin{table}[h]
\caption{Carreau-Yasuda parameters, density, and surface tension as a function of the alginate concentration $c$ at 25$^\circ$C.}
\centering
%\begin{center}
{\footnotesize
\begin{tabular}{lccccccc}
\hline
$c$[\%] & $\eta_0$[Pa s] & $\eta_{\infty}$[Pa s] & $\lambda_{CY}$[s] & $a$ & $n$ & $\rho$[kg/m$^3$] & $\sigma$[mN/m] \\
\hline
0.1 & 0.001 & 0.001 & 0 & 2 & 1 & 998 & 72\\ %[-0.4cm]
0.2 & 0.003 & 0.001 & 0.001 & 2 & 1 & 998 & 72\\ %[-0.4cm]
0.3 & 0.004 & 0.001 & 0.002 & 2 & 1 & 998 & 72\\ %[-0.4cm]
0.4 & 0.005 & 0.001 & 0.002 & 2 & 1 & 999 & 72\\ %[-0.4cm]
0.5 & 0.005 & 0.001 & 0.003 & 2 & 1 & 999 & 72\\ %[-0.4cm]
0.6 & 0.006 & 0.001 & 0.003 & 2 & 0.95 & 1000 & 72\\ %[-0.4cm]
0.7 & 0.007 & 0.001 & 0.003 & 1.925 & 0.936 & 1000 & 72\\ %[-0.4cm]
0.8 & 0.011 & 0.001 & 0.004 & 1.85 & 0.923 & 1001 & 72\\ %[-0.4cm]
0.9 & 0.012 & 0.001 & 0.004 & 1.775 & 0.909 & 1003 & 71.9\\ %[-0.4cm]
1 & 0.014 & 0.001 & 0.005 & 1.7 & 0.895 & 1004 & 71.8\\ %[-0.4cm]
1.15 & 0.016 & 0.001 & 0.005 & 1.587 & 0.87 & 1005 & 71.7\\ %[-0.4cm]
1.5 & 0.036 & 0.001 & 0.005 & 1.48 & 0.823 & 1006 & 71.5\\ %[-0.4cm]
1.75 & 0.04 & 0.001 & 0.006 & 1.363 & 0.786 & 1007 & 71.3\\ %[-0.4cm]
2 & 0.057 & 0.001 & 0.006 & 1.25 & 0.75 & 1008 & 71.1\\ %[-0.4cm]
2.25 & 0.072 & 0.001 & 0.006 & 1.15 & 0.713 & 1009 & 70.8\\ %[-0.4cm]
2.5 & 0.131 & 0.001 & 0.006 & 1.05 & 0.675 & 1010 & 70.5\\ %[-0.4cm]
2.75 & 0.137 & 0.001 & 0.006 & 0.95 & 0.637 & 1011 & 70.1\\ %[-0.4cm]
3 & 0.205 & 0.001 & 0.006 & 0.85 & 0.6 & 1012 & 69.7\\ %[-0.4cm]
3.5 & 0.32 & 0.001 & 0.013 & 0.735 & 0.585 & 1014 & 68.9\\ %[-0.4cm]
4 & 0.69 & 0.001 & 0.02 & 0.62 & 0.57 & 1016 & 68\\ %[-0.4cm]
4.5 & 0.7 & 0.001 & 0.023 & 0.58 & 0.6 & 1018 & 66.5\\ %[-0.4cm]
5 & 1.75 & 0.001 & 0.025 & 0.5 & 0.49 & 1020 & 65\\ %[-0.4cm]
5.5 & 1.75 & 0.001 & 0.025 & 0.5 & 0.495 & 1022 & 63.3\\ %[-0.4cm]
6 & 3.45 & 0.001 & 0.025 & 0.47 & 0.43 & 1024 & 61.5\\ %[-0.4cm]
6.5 & 3.78 & 0.001 & 0.045 & 0.48 & 1 & 1026 & 60\\ %[-0.4cm]
7 & 7 & 0.001 & 0.065 & 0.49 & 0.45 & 1028 & 58.5\\ %[-0.4cm]
8 & 15 & 0.001 & 0.17 & 0.55 & 0.46 & 1032 & 57.8\\ %[-0.4cm]
9 & 32 & 0.001 & 0.25 & 0.5 & 0.41 & 1036 & 57.8\\ %[-0.4cm]
%10 & 61 & 0.001 & 0.4 & 0.5 & 0.4 & 1040 & 57.8\\ %[-0.4cm]
\hline
\end{tabular} }   
%\end{center}
\label{tb:table-1}
\end{table}  
      
Initial upper droplet diameters $d_0$ are in the range of 1.5mm-2.6mm. Additionally, the nozzle diameter $D$ ranges from 1.35mm to 1.85mm, and the pool's dimensions (diameter and height) $L$ are kept fixed at 4.35mm. Lastly, the distance $H$ between the bottom of the pool and the top of the lower droplet is kept at 5.1mm for all flow cases explored here. A movie showing a standard experiment is available on {\color{purple}{https://www.youtube.com/watch?v=ctmVaL1ZbV8}}.

The dimensionless numbers governing the problem, derived from the Buckingham-$\Pi$ theorem, are based on the dimensional variables $d_0$, $\rho$, $\eta_0$, $\eta_{\infty}$, $\lambda_e$, $\sigma$, $g$ (the $z$ component of the gravity vector), $U_0$ (liquid mean vertical velocity in the nozzle), $\rho_{air}$ and $\eta_{air}$, with the fundamental units of mass [kg], distance [m] and time [s]. These variables lead to seven important dimensionless quantities:
\begin{equation}
\Pi_1 = \frac{\eta_0 \sqrt{\sigma/(\rho d_0^3)}}{\sigma/d_0}  \, ,
\label{eq2.1}
\end{equation}
\begin{equation}
\Pi_2 = \lambda_e \sqrt{\sigma/(\rho d_0^3)}  \, ,
\label{eq2.2}
\end{equation}
\begin{equation}
\Pi_3 = \frac{\rho g d_0}{\sigma/d_0}  \, ,
\label{eq2.3}
\end{equation}
\begin{equation}
\Pi_4 = \frac{\rho U_0^2}{\sigma/d_0}  \, ,
\label{eq2.4}
\end{equation}
\begin{equation}
\Pi_5 = \frac{\eta_{\infty} \sqrt{\sigma/(\rho d_0^3)}}{\sigma/d_0}  \, ,
\label{eq2.5}
\end{equation}
\begin{equation}
\Pi_6 = \frac{\eta_{air} \sqrt{\sigma/(\rho d_0^3)}}{\sigma/d_0}   \, ,
\label{eq2.6}
\end{equation}
\begin{equation}
\Pi_7 =  \frac{\rho_{air}}{\rho}   \, .
\label{eq2.7}
\end{equation}
$\Pi_1$ represents the $\eta_0$-based Ohnesorge number $\mathrm{Oh_0}$, $\Pi_2$ is the intrinsic Deborah number $\mathrm{De}$, $\Pi_3$ corresponds to the Bond number $\mathrm{Bo}$, $\Pi_4$ is the Weber number based on the solution's mean velocity into the nozzle $\mathrm{We_{nozzle}}$, $\Pi_5$ is the $\eta_{\infty}$-based Ohnesorge number $\mathrm{Oh_{\infty}}$, $\Pi_6$ is the air-based Ohnesorge number $\mathrm{Oh_{air}}$, $\Pi_7$ is the density ratio, and $\sqrt{\sigma/(\rho d_0^3)}$ is a characteristic strain rate associated with the capillary-thinning process. In the present work, $\mathrm{Bo} < 1.4$,
${\mathrm{Oh_{\infty}}} \sim 10^{-3}$, ${\mathrm{Oh_{air}}} \lesssim 10^{-5}$, $\rho_{air}/\rho \lesssim 10^{-3}$, and $\mathrm{We_{nozzle}} \lesssim 10^{-4}$ [$U_0$ is much lower than the characteristic stretching velocity $\sqrt{\sigma/(\rho d_0)}$]. Thus, the flow scenarios considered here vary with only two key dimensionless numbers:  
\begin{equation}
\mathrm{Oh_0} = \frac{\eta_0 \sqrt{\sigma/(\rho d_0^3)}}{\sigma/d_0}    \, ,
\label{eq2.8}
\end{equation}
\begin{equation}
\mathrm{De} = \lambda_e \sqrt{\sigma/(\rho d_0^3)}    \, .
\label{eq2.9}
\end{equation}
The $\eta_0$-based Ohnesorge number ranges from 0.002 to 78, while the Deborah number varies from 0.1 to 491 (the rheometric protocol leading to the relaxation time values used for the calculation of the Deborah numbers will be detailed in the following pages).  

%%%%%%%%%%%%%%%%%%%%%%%%%%%%%%%%%%%%%%%%%%%%%%%%%%%%%%%%%%%%%%%%%%%%%%%%%%%%%%
%%%%%%%%%%%%%%%%%%%%%%%%%%%%%%%%%%%%%%%%%%%%%%%%%%%%%%%%%%%%%%%%%%%%%%%%%%%%%%
\section{Results and Discussion} \label{RD}
%%%%%%%%%%%%%%%%%%%%%%%%%%%%%%%%%%%%%%%%%%%%%%%%%%%%%%%%%%%%%%%%%%%%%%%%%%%%%%
%%%%%%%%%%%%%%%%%%%%%%%%%%%%%%%%%%%%%%%%%%%%%%%%%%%%%%%%%%%%%%%%%%%%%%%%%%%%%%

\begin{figure*}[htp]
\centering    
\includegraphics[width=1\linewidth]{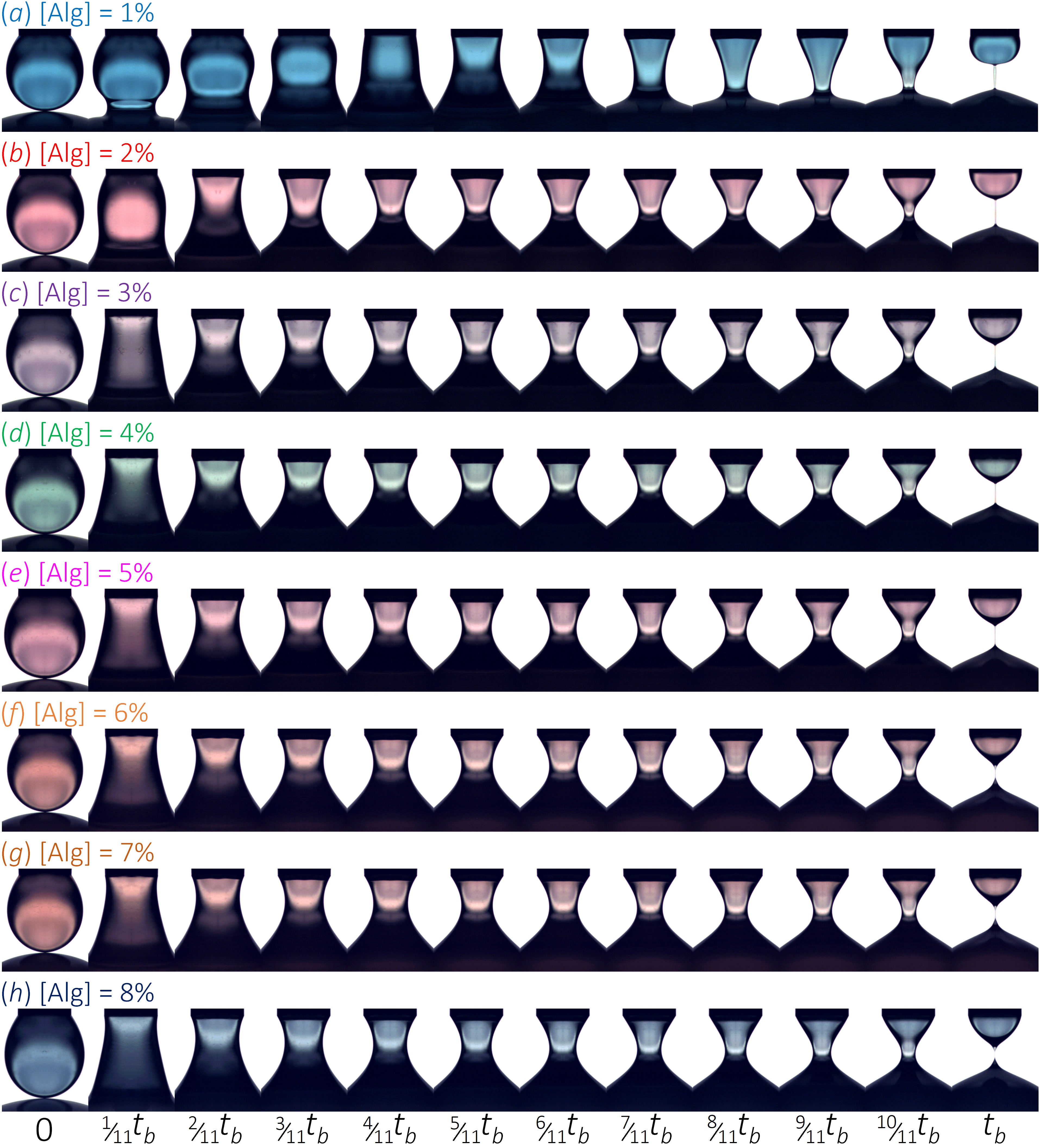}
\vspace{-0.7cm}    
\caption{{\color{darkgray}{Snapshots illustrating the droplets' coalescence followed by the capillary thinning and breakup of the formed filaments at eight alginate concentrations by weight $c$: (\textit{a}) 1\%; (\textit{b}) 2\%; (\textit{c}) 3\%; (\textit{d}) 4\%; (\textit{e}) 5\%; (\textit{f}) 6\%; (\textit{g}) 7\%; and (\textit{h}) 8\%. The initial droplet diameter is approximately 2.1mm in all shown cases. Nozzle diameter (black upper rectangle) $D$ = 1.37mm. The time interval between subsequent snapshots is $\frac{1}{11}t_b$.}}} 
\vspace{-0.2cm} 
\label{fig-2}
\end{figure*}

\begin{figure*}[htp]
\centering    
\includegraphics[width=1\linewidth]{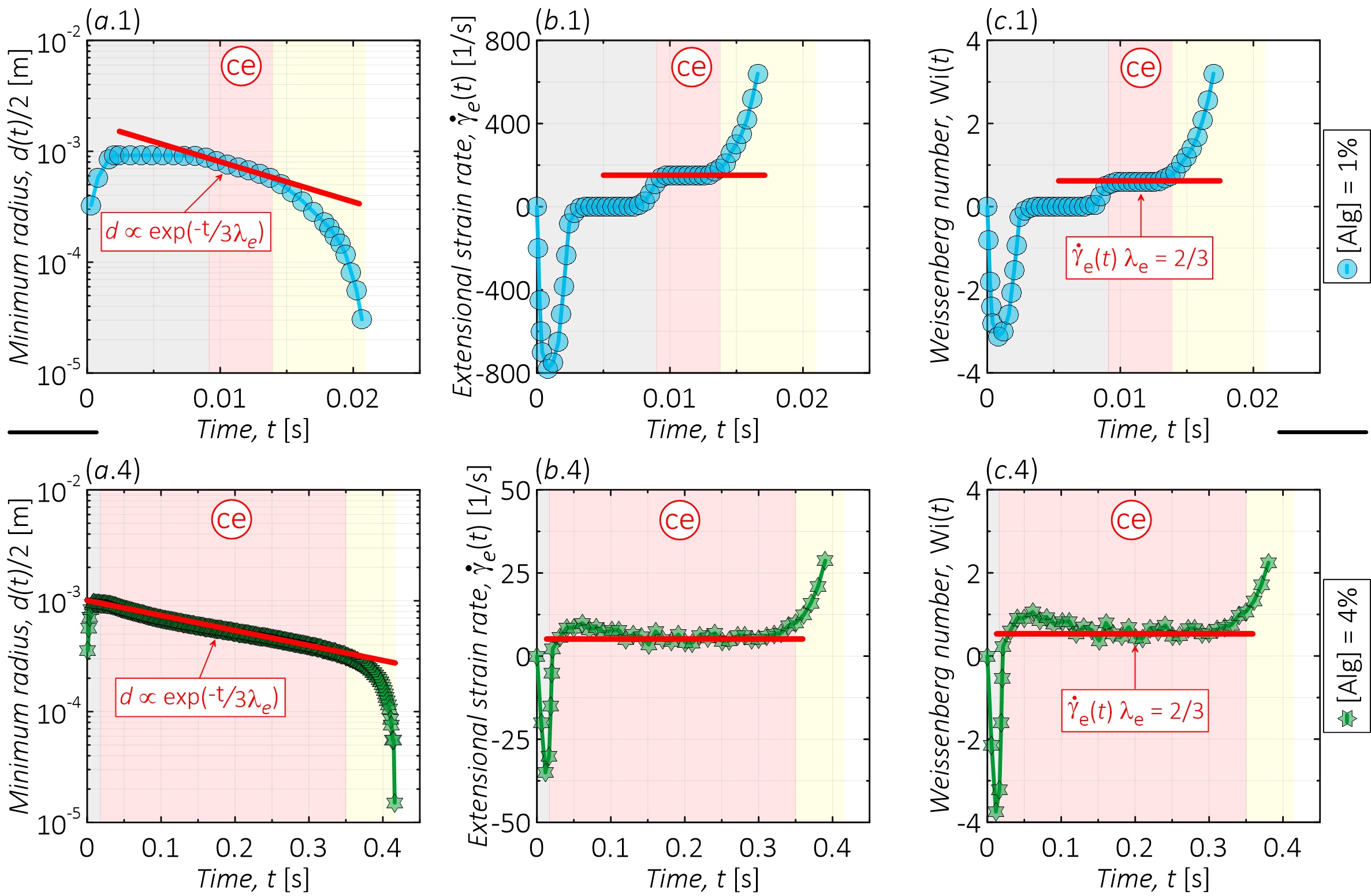}
\vspace{-0.7cm}    
\caption{{\color{darkgray}{(\textit{a}) Instantaneous filament minimum radius $d(t)/2$ against time $t$ at $c =$ 1\% (\textit{a}.1), and $c =$ 4\% (\textit{a}.4). (\textit{b}) Instantaneous extensional strain rate ${\dot{\gamma}}_e(t)$ against time $t$ at $c =$ 1\% (\textit{b}.1), and $c =$ 4\% (\textit{b}.4). (\textit{c}) Instantaneous Weissenberg number $\mathrm{Wi}(t)$ against time $t$ at $c =$ 1\% (\textit{c}.1), and $c =$ 4\% (\textit{c}.4). The grey boxes denote the initial droplet coalescence and formation of the liquid bridge. The red boxes highlight the capillary-elastic filament thinning. The yellow boxes indicate the final thinning stage leading to rupture. The error bars are comparable in size to the symbols.}}} 
\vspace{-0.2cm} 
\label{fig-3}
\end{figure*}

Sub-figures \ref{fig-2}(\textit{a})-\ref{fig-2}(\textit{h}) illustrate droplets' coalescence followed by the capillary thinning and breakup of the formed filaments at eight alginate concentrations by weight $c$: 1\%, 2\%, 3\%, 4\%, 5\%, 6\%, 7\%, and 8\%, respectively. The initial droplet diameter is approximately 2.1mm in all shown cases. Since the droplets' density slightly varies from 1004kg/m$^3$ to 1032kg/m$^3$ as the polymer concentration ranges from 1\% to 8\% (see table \ref{tb:table-1}), figure \ref{fig-2} primarily stresses viscoelastic effects on the capillary thinning (e.g., the Bond number is practically constant, whereas $\mathrm{Oh}$ and $\mathrm{De}$ augment with the increasing $c$). The sub-figures are composed of 12 successive alginate-in-water snapshots that show the time evolution of the thinning process until the filaments' breakup. Note that only the centre free-surface part of the flow is depicted. The time interval between subsequent snapshots is $\frac{1}{11}t_b$.

At $c$ = 1\%, the filament diameter evolves in a non-uniform way along the $z$ direction, leading to a fluid conical shape ($\frac{7}{11}t_b \leq t \leq \frac{10}{11}t_b$) that turns into an unchanging filament in the late stage of the thinning process ($\frac{10}{11}t_b < t < t_b$). The conical form is observed in the present work for $c\leq$ 1\%, and it is very often linked with relevant inertial effects within the capillary bridge \citep{Khoury_2026}. Nevertheless, such a heterogeneity tends to vanish as the alginate concentration increases, thereby amplifying the viscoelastic stresses relative to the inertial and capillary stresses, and leading to a homogeneous centre filament that uniformly thins until it breaks up. Logically, higher viscoelasticity delays the growth of perturbations at the liquid/air interface, favouring the formation of a homogeneous liquid bridge between the upper and lower droplets. Uniform centre filaments are typically associated with both viscous and elastic effects \citep[$\mathrm{Oh} \gtrsim 1$, and $\mathrm{De} \gtrsim 1$;][]{McKinley_2000, Anna_2001, Khoury_2026}.

The instantaneous minimum radius $d(t)/2$ of the filaments at $c =$ 1\% and $c =$ 4\% are respectively plotted against time $t$ in sub-figures \ref{fig-3}(\textit{a}.1) and \ref{fig-3}(\textit{a}.4). The initial radius augmentation observed in the curves (and highlighted by the grey boxes) results from the droplets coalescence and formation of the liquid bridge, which is followed by the surface-tension-induced filament thinning. In the latter stage, a non-linear radius decay is clearly observed in both cases. The decreasing radius profile can be divided into an exponential region (red boxes) and a subsequent rapid decay (yellow boxes), leading to rupture. The exponential segment of the curves is fitted by equation \ref{eq:intro-1}, as illustrated by the red lines, which leads to an extensional relaxation time $\lambda_e =$ 4ms at $c =$ 1\%, and $\lambda_e =$ 110ms at $c =$ 4\%. By combining equation \ref{eq:intro-1} with the instantaneous uniaxial extensional strain rate ${\dot{\gamma}}_e(t)$ 
\begin{equation}
{\dot{\gamma}}_e(t) = -\frac{2}{d(t)} \frac{\mathrm{d}d(t)}{\mathrm{d} t}  \, ,
\label{eq:gammadote}
\end{equation}
one finds that, in the capillary-elastic regime, ${\dot{\gamma}}_e$ must be constant. This is confirmed by sub-figures \ref{fig-3}(\textit{b}.1) and \ref{fig-3}(\textit{b}.4), where ${\dot{\gamma}}_e$ is plotted against time $t$ for filaments at $c =$ 1\% and $c =$ 4\%, respectively. The constant plateaus are highlighted by the horizontal straight lines (in the red zones), e.g., ${\dot{\gamma}}_e(t) \approx 167$s$^{-1}$ at $c =$ 1\%, and ${\dot{\gamma}}_e(t) \approx 6$s$^{-1}$ at $c =$ 4\%. Furthermore, in the capillary-elastic segment, the instantaneous Weissenberg number defined as $\mathrm{Wi}(t) = \lambda_e {\dot{\gamma}}_e(t)$ must satisfy         
\begin{equation}
\mathrm{Wi}(t) = \lambda_e {\dot{\gamma}}_e(t) = \frac{2}{3}  \, .
\label{eq:gammadote}
\end{equation}
Indeed, by multiplying the instantaneous uniaxial extensional strain rate given in sub-figures \ref{fig-3}(\textit{b}.1) and \ref{fig-3}(\textit{b}.4) by the extensional relaxation time obtained by fitting the exponential segment of the curves displayed in sub-figures \ref{fig-3}(\textit{a}.1) and \ref{fig-3}(\textit{a}.4) with equation \ref{eq:intro-1}, one finds that $\mathrm{Wi}(t) \approx 2/3$ during the exponential radius decay, as confirmed by sub-figures \ref{fig-3}(\textit{c}.1) and \ref{fig-3}(\textit{c}.4) where instantaneous Weissenberg number is plotted against time. It is also crucial to emphasise that the time interval $\Delta t$ during which ${\dot{\gamma}}_e$ is approximately constant and $\mathrm{Wi}(t) \approx 2/3$ is an increasing function of the polymer concentration. At $c =$ 1\%, for instance, $\Delta t$ corresponds to approximately 6\% of the breakup time ($\Delta t \approx 0.003$s and $t_b \approx 0.021$s), raising up to  75\% of $t_b$ at $c =$ 4\% ($\Delta t \approx 0.3$s and $t_b \approx 0.41$s). This tendency is also observed at all the used polymer concentrations (not shown for brevity). In other words, as alginate is added to the solution, the filament thinning becomes primarily dominated by the competition between the capillary pressure and the elastic stress. In contrast, the thinning dynamics tend to be driven by capillary-inertial balance as the alginate concentration vanishes and the polymeric solution approaches the solvent \citep[see][]{Khoury_2026}.     
   
\begin{figure}[htp]
\centering    
\includegraphics[width=1\linewidth]{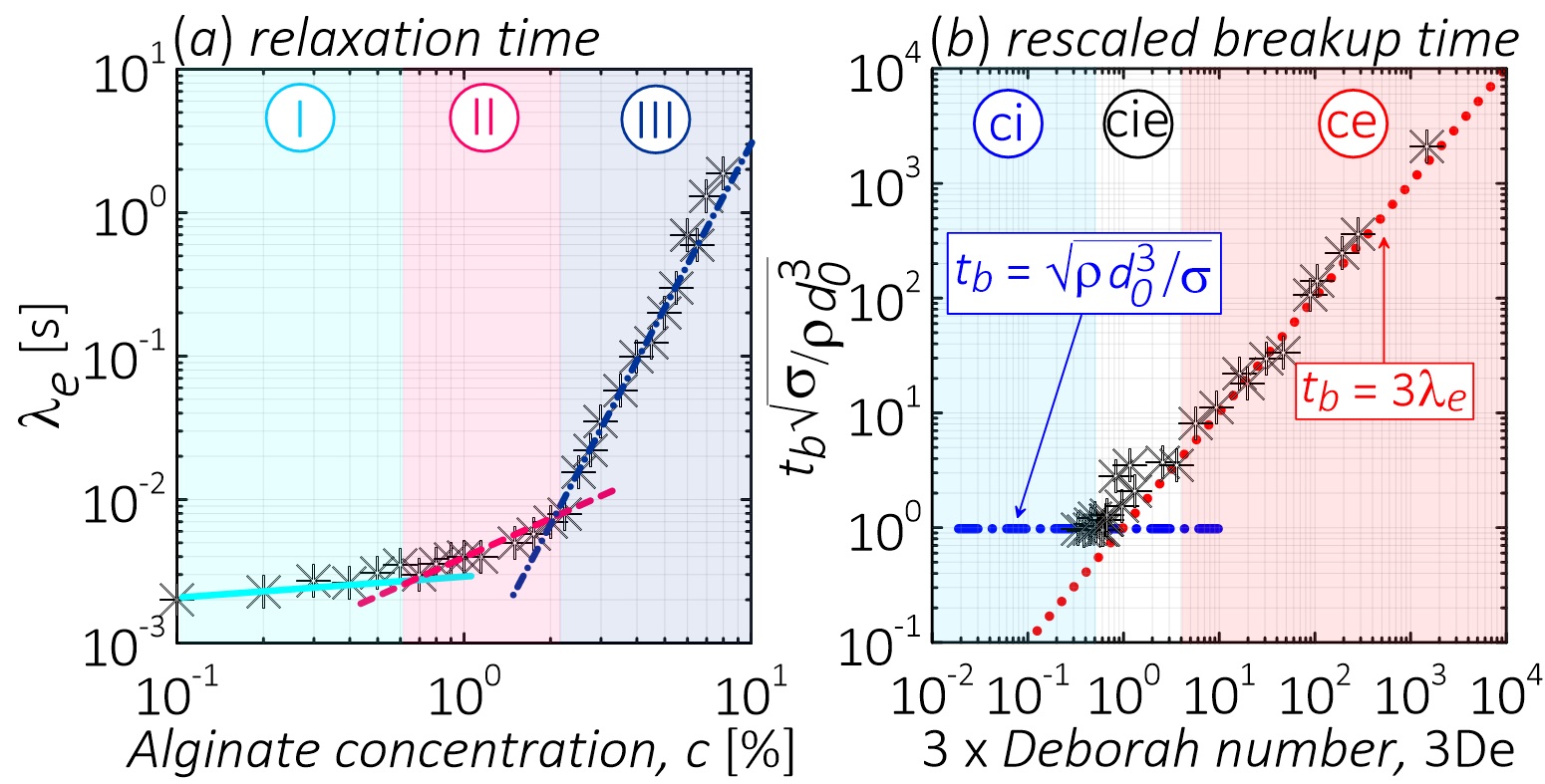}
\vspace{-0.7cm}    
\caption{{\color{darkgray}{(\textit{a})Extensional relaxation time $\lambda_e$ as a function of the alginate concentration $c$. The data are divided into three zones representing concentration regimes: diluted (in cyan; I), semi-diluted unentangled (in pink; II), and semi-diluted entangled (in navy blue; III). In each zone, the data are fitted by a power-law function $\lambda_e \propto c^{\beta}$ whose exponents are: $\beta = 1/2$ in I (cyan solid line), $\beta = 1$ in II (pink dashed line), and $\beta = 3.75$ in III (navy blue dash-dotted line). (\textit{b}) Rescaled breakup time $t_b/\sqrt{\rho d_0^3/\sigma}$ against three times the Deborah number $3\mathrm{De}$. The results collapse across a master curve divided into three regions highlighting the important capillary-thinning scenarios: capillary-inertial (ci; in blue), for which $t_b = \sqrt{\rho d_0^3/\sigma}$; capillary-elastic (ce; in red), where $t_b = 3\lambda_e$; and a mixed regime (cie; in white), for which capillary, inertial and elastic effects all play an essential role. The mixed regime (transitioning zone) emerges when $ 0.5 < \mathrm{De} < 4$. The error bars are comparable in size to the symbols.}}} 
\vspace{-0.2cm} 
\label{fig-4}
\end{figure}

By applying the procedure illustrated in figure \ref{fig-3} to the solutions considered here, we plot in sub-figure \ref{fig-4}(\textit{a}) the extensional relaxation time $\lambda_e$ as a function of the alginate concentration. Similar to $\eta_0$, the data are categorised into three regions related to the concentration regimes: diluted for $c \le 0.65$\% (I; cyan region), semi-diluted unentangled for $0.65$\% $< c < 2.2$\% (II; pink region), and semi-diluted entangled for $c \geq 2.2$\% (III; navy-blue region). In each concentration regime, the extensional relaxation time can be fitted by a power-law equation $\lambda_e \propto c^{\beta}$ (as pointed out by the straight lines) whose exponent magnifies when moving from the diluted towards the semi-diluted entangled concentration regime, as a result of the polymer chains' interpenetration: $\beta =$ 1/2 (cyan solid line), 1 (pink dashed line), and 3.75 (navy blue dash-dotted line) in the diluted, semi-diluted unentangled, and semi-diluted entangled concentration regime, respectively. %Curiously, these values are the same ones used to fit the data in sub-figure \ref{fig-3}(\textit{d}).

Lastly, based on the findings highlighted in figure \ref{fig-3}, we propose scaling laws for $t_b$ by considering the capillary-inertial regime appearing as the alginate concentration becomes less pronounced, and the capillary-elastic regime resulting from an augmenting alginate amount. In the capillary-inertial thinning regime, filament thinning emerges from a balance between the capillary pressure ($\sim \sigma/ d_0$) and inertial stresses ($\sim \rho u_c^2$, where $u_c$ is the characteristic filament thinning velocity approximated as $u_c \sim d_0/t_b$), leading to
\begin{equation}
t_{ci} = \sqrt{\frac{\rho d_0^3}{\sigma}} ~~~ \text{(capillary-inertial time)} \, ,
\label{eq.tb-ci}
\end{equation}
a well-known characteristic capillary-inertial time extensively discussed in previous works \citep{Peregrine_1990, Eggers_1993, Eggers_2005, Javadi_2013, Deblais_2018, Dinic_2019, Bazazi_2025}. On the other hand, in the capillary-elastic regime, the capillary pressure is mainly counter-balanced by elastic stresses during the filament's uniaxial extensional deformation, which leads to equation \ref{eq:intro-1}. According to the latter, the characteristic capillary-elastic time is 
\begin{equation}
t_{ce} = 3\lambda_e ~~~ \text{(capillary-elastic time)} \, .
\label{eq.tb-ce}
\end{equation}
Hence, for the capillary-inertial regime, $t_b$ is expected to scale with $\sqrt{\rho d_0^3/\sigma}$ (also called the Rayleigh time scale), while in the capillary-elastic regime, $t_b$ should scale with $\lambda_e$. By equating the above characteristic times (equations \ref{eq.tb-ci} and \ref{eq.tb-ce}), we find $3\mathrm{De} = 1$. More specifically, $3\mathrm{De}$ tends to 1 when transitioning from the capillary-inertial thinning to the capillary-elastic one. This theoretical prediction is corroborated by sub-figure \ref{fig-4}(\textit{b}), in which the rescaled breakup time $t_b \sqrt{\sigma/\rho d_0^3}$ (e.g., the breakup time made dimensionless by the characteristic capillary-inertial breakup time given by equation \ref{eq.tb-ci}) is plotted as a function of three times the Deborah number for all the alginate solutions considered here. The results collapse across a master curve divided into three regions highlighting the important capillary-thinning scenarios: capillary-inertial (ci; in blue), for which $t_b = \sqrt{\rho d_0^3/\sigma}$; capillary-elastic (ce; in red), where $t_b = 3\lambda_e$; and a mixed regime (cie; in white), for which capillary, inertial and elastic effects all play an essential role. As underlined by our theoretical analyses, a mixed regime (transitioning zone) emerges when $ 0.5 < \mathrm{De} < 4$. Note as well that the intersection between the straight lines representing the scaling laws occurs at $3\mathrm{De} = 1$, as theoretically predicted. Finally, it is crucial to emphasise that the master curve revealed by sub-figure \ref{fig-4}(\textit{b}) can be used to estimate either $\sigma$ or $\lambda_e$ exclusively based on $t_b$. In other words, the master curve can be viewed as either a tensiometer or a uniaxial extensional rheometer.   
 
%%%%%%%%%%%%%%%%%%%%%%%%%%%%%%%%%%%%%%%%%%%%%%%%%%%%%%%%%%%%%%%%%%%%%%%%%%%%%%
%%%%%%%%%%%%%%%%%%%%%%%%%%%%%%%%%%%%%%%%%%%%%%%%%%%%%%%%%%%%%%%%%%%%%%%%%%%%%%
\section{Concluding Remarks}
%%%%%%%%%%%%%%%%%%%%%%%%%%%%%%%%%%%%%%%%%%%%%%%%%%%%%%%%%%%%%%%%%%%%%%%%%%%%%%
%%%%%%%%%%%%%%%%%%%%%%%%%%%%%%%%%%%%%%%%%%%%%%%%%%%%%%%%%%%%%%%%%%%%%%%%%%%%%%
%\vspace*{-0.5cm}

We have presented here a mixed experimental and theoretical study allowing to assess the extensional relaxation time of alginate solutions by using dripping-onto-droplet capillary breakup rheometry (DoD), e.g., the capillary thinning and breakup of fluid filaments formed following the coalescence of a millimetric-nozzle-generated pendant drop with a lower droplet cap of the same fluid contained in a millimetric pool in ambient air. The experiments were recorded with a high-speed camera using alginate solutions with alginate concentrations ranging from 0.1\% to 9\% by weight in deionised water. The results were depicted by considering the dynamics of fluid filament thinning, stress balances, and scaling laws. 

The solutions' extensional relaxation times were obtained by fitting the filaments' diameter decay with the equation derived by Entov and Hinch \citep{Entov_1997} for an Oldroyd-B/Hookean dumbbell, according to which $d(t) \propto \exp{\left( -t/3\lambda_e \right)}$. Extensional relaxation times of about 2ms were found for the diluted alginate solutions. Additionally, three flow regimes were highlighted: the capillary-inertial, resulting from a competition between capillary and inertial stresses, for which the breakup time $t_b$ scales with $\sqrt{\rho d_0^3/\sigma}$;  the capillary-elastic regime, emerging from a balance between capillary and elastic stresses, for which $t_b \sim \lambda_e$; and the mixed capillary-inertial-elastic for which capillarity, inertia and elasticity are all relevant. More importantly, we have also shown that $\lambda_e$ can be directly extracted from the filament breakup time through a master curve connecting $t_b$ to the intrinsic Deborah number defined as $\mathrm{De} = \lambda_e \sqrt{\sigma/(\rho d_0^3)}$.

Finally, it would be interesting to extend these analyses in future work by considering a wider selection of polymers. Another interesting possibility would be using DoD in a three-phase-flow configuration to study, for instance, mass-transfer and/or ionic crosslinking effects on the capillary-thinning process, i.e., the upper polymer-based droplet undergoes ion-induced gelation following its coalescence with the lower droplet containing ions \citep[the polymer molecules crosslink as the ions diffuse within the fluid filament;][]{Godefroid_2025}.

%%%%%%%%%%%%%%%%%%%%%%%%%%%%%%%%%%%%%%%%%%%%%%%%%%%%%%%%%%%%%%%%%%%%%%%%%%%%%%
%%%%%%%%%%%%%%%%%%%%%%%%%%%%%%%%%%%%%%%%%%%%%%%%%%%%%%%%%%%%%%%%%%%%%%%%%%%%%%
\textit{\textbf{Acknowledgements:}} 
%%%%%%%%%%%%%%%%%%%%%%%%%%%%%%%%%%%%%%%%%%%%%%%%%%%%%%%%%%%%%%%%%%%%%%%%%%%%%%
%%%%%%%%%%%%%%%%%%%%%%%%%%%%%%%%%%%%%%%%%%%%%%%%%%%%%%%%%%%%%%%%%%%%%%%%%%%%%%
This research is co-funded by The Transition Institute 1.5 of Mines Paris - PSL. The authors would also like to acknowledge the support of the PSL Research University under the program `Investissements d'Avenir' launched by the French Government and implemented by the French National Research Agency (ANR) with the reference ANR-10-IDEX-0001-02 PSL, the ANR (ANR-24-CE06-0591) and the S\~ao Paulo Research Foundation (2024/04769-1) for supporting the GENNIAL project under the PRCI program (`Projet de Recherche Collaborative Internationale'), and the Metasoft Matter program for co-funding this research.    

%\newpage
%\clearpage

%%%%%%%%%%%%%%%%%%%%%%%%%%%%%%%%%%%%%%%%%%%%%%%%%%%%%%%%%%%%%%%%%%%%%%%%%%%%%%
%%%%%%%%%%%%%%%%%%%%%%%%%%%%%%%%%%%%%%%%%%%%%%%%%%%%%%%%%%%%%%%%%%%%%%%%%%%%%%
%\bibliographystyle{./apsrev4-2}
\bibliography{DoD-ref}% Produces the bibliography via BibTeX.

\begin{thebibliography}{64}%
\makeatletter
\providecommand \@ifxundefined [1]{%
 \@ifx{#1\undefined}
}%
\providecommand \@ifnum [1]{%
 \ifnum #1\expandafter \@firstoftwo
 \else \expandafter \@secondoftwo
 \fi
}%
\providecommand \@ifx [1]{%
 \ifx #1\expandafter \@firstoftwo
 \else \expandafter \@secondoftwo
 \fi
}%
\providecommand \natexlab [1]{#1}%
\providecommand \enquote  [1]{``#1''}%
\providecommand \bibnamefont  [1]{#1}%
\providecommand \bibfnamefont [1]{#1}%
\providecommand \citenamefont [1]{#1}%
\providecommand \href@noop [0]{\@secondoftwo}%
\providecommand \href [0]{\begingroup \@sanitize@url \@href}%
\providecommand \@href[1]{\@@startlink{#1}\@@href}%
\providecommand \@@href[1]{\endgroup#1\@@endlink}%
\providecommand \@sanitize@url [0]{\catcode `\\12\catcode `\$12\catcode
  `\&12\catcode `\#12\catcode `\^12\catcode `\_12\catcode `\%12\relax}%
\providecommand \@@startlink[1]{}%
\providecommand \@@endlink[0]{}%
\providecommand \url  [0]{\begingroup\@sanitize@url \@url }%
\providecommand \@url [1]{\endgroup\@href {#1}{\urlprefix }}%
\providecommand \urlprefix  [0]{URL }%
\providecommand \Eprint [0]{\href }%
\providecommand \doibase [0]{https://doi.org/}%
\providecommand \selectlanguage [0]{\@gobble}%
\providecommand \bibinfo  [0]{\@secondoftwo}%
\providecommand \bibfield  [0]{\@secondoftwo}%
\providecommand \translation [1]{[#1]}%
\providecommand \BibitemOpen [0]{}%
\providecommand \bibitemStop [0]{}%
\providecommand \bibitemNoStop [0]{.\EOS\space}%
\providecommand \EOS [0]{\spacefactor3000\relax}%
\providecommand \BibitemShut  [1]{\csname bibitem#1\endcsname}%
\let\auto@bib@innerbib\@empty
%</preamble>
\bibitem [{\citenamefont {Lee}\ and\ \citenamefont {Mooney}(2012)}]{Lee_2012}%
  \BibitemOpen
  \bibfield  {author} {\bibinfo {author} {\bibfnamefont {K.~Y.}\ \bibnamefont
  {Lee}}\ and\ \bibinfo {author} {\bibfnamefont {D.~J.}\ \bibnamefont
  {Mooney}},\ }\bibfield  {title} {\bibinfo {title} {Alginate: properties and
  biomedical applications},\ }\href@noop {} {\bibfield  {journal} {\bibinfo
  {journal} {Progress in Polymer Science}\ }\textbf {\bibinfo {volume} {37}},\
  \bibinfo {pages} {106} (\bibinfo {year} {2012})}\BibitemShut {NoStop}%
\bibitem [{\citenamefont {Sun}\ \emph {et~al.}(2013)\citenamefont {Sun},
  \citenamefont {Tan}, \citenamefont {Sun},\ and\ \citenamefont
  {Tan}}]{Sun_2013}%
  \BibitemOpen
  \bibfield  {author} {\bibinfo {author} {\bibfnamefont {J.}~\bibnamefont
  {Sun}}, \bibinfo {author} {\bibfnamefont {H.}~\bibnamefont {Tan}}, \bibinfo
  {author} {\bibfnamefont {J.}~\bibnamefont {Sun}},\ and\ \bibinfo {author}
  {\bibfnamefont {H.}~\bibnamefont {Tan}},\ }\bibfield  {title} {\bibinfo
  {title} {Alginate-based biomaterials for regenerative medicine
  applications},\ }\href@noop {} {\bibfield  {journal} {\bibinfo  {journal}
  {Materials}\ }\textbf {\bibinfo {volume} {6}},\ \bibinfo {pages} {1285}
  (\bibinfo {year} {2013})}\BibitemShut {NoStop}%
\bibitem [{\citenamefont {Vicini}\ \emph {et~al.}(2015)\citenamefont {Vicini},
  \citenamefont {Castellano}, \citenamefont {Mauri},\ and\ \citenamefont
  {Marsano}}]{Vicini_2015}%
  \BibitemOpen
  \bibfield  {author} {\bibinfo {author} {\bibfnamefont {S.}~\bibnamefont
  {Vicini}}, \bibinfo {author} {\bibfnamefont {M.}~\bibnamefont {Castellano}},
  \bibinfo {author} {\bibfnamefont {M.}~\bibnamefont {Mauri}},\ and\ \bibinfo
  {author} {\bibfnamefont {E.}~\bibnamefont {Marsano}},\ }\bibfield  {title}
  {\bibinfo {title} {Gelling process for sodium alginate: new technical
  approach by using calcium rich microspheres},\ }\href@noop {} {\bibfield
  {journal} {\bibinfo  {journal} {Carbohydrate Polymers}\ }\textbf {\bibinfo
  {volume} {134}},\ \bibinfo {pages} {767} (\bibinfo {year}
  {2015})}\BibitemShut {NoStop}%
\bibitem [{\citenamefont {Castellano}\ \emph {et~al.}(2019)\citenamefont
  {Castellano}, \citenamefont {Darawish}, \citenamefont {Dodero},\ and\
  \citenamefont {Vicini}}]{Castellano_2019}%
  \BibitemOpen
  \bibfield  {author} {\bibinfo {author} {\bibfnamefont {M.}~\bibnamefont
  {Castellano}}, \bibinfo {author} {\bibfnamefont {M.~A.~R.}\ \bibnamefont
  {Darawish}}, \bibinfo {author} {\bibfnamefont {A.}~\bibnamefont {Dodero}},\
  and\ \bibinfo {author} {\bibfnamefont {S.}~\bibnamefont {Vicini}},\
  }\bibfield  {title} {\bibinfo {title} {Electrospun composite mats of alginate
  with embedded silver nanoparticles},\ }\href@noop {} {\bibfield  {journal}
  {\bibinfo  {journal} {Journal of Thermal Analysis and Calorimetry}\ }\textbf
  {\bibinfo {volume} {137}},\ \bibinfo {pages} {767–778} (\bibinfo {year}
  {2019})}\BibitemShut {NoStop}%
\bibitem [{\citenamefont {Dodero}\ \emph {et~al.}(2019)\citenamefont {Dodero},
  \citenamefont {Vicini}, \citenamefont {Alloisio},\ and\ \citenamefont
  {Castellano}}]{Dodero_2019}%
  \BibitemOpen
  \bibfield  {author} {\bibinfo {author} {\bibfnamefont {A.}~\bibnamefont
  {Dodero}}, \bibinfo {author} {\bibfnamefont {S.}~\bibnamefont {Vicini}},
  \bibinfo {author} {\bibfnamefont {M.}~\bibnamefont {Alloisio}},\ and\
  \bibinfo {author} {\bibfnamefont {M.}~\bibnamefont {Castellano}},\ }\bibfield
   {title} {\bibinfo {title} {Sodium alginate solutions: correlation between
  rheological properties and spinnability},\ }\href@noop {} {\bibfield
  {journal} {\bibinfo  {journal} {Journal of Materials Science - Polymers}\
  }\textbf {\bibinfo {volume} {54}},\ \bibinfo {pages} {8034} (\bibinfo {year}
  {2019})}\BibitemShut {NoStop}%
\bibitem [{\citenamefont {Dodero}\ \emph {et~al.}(2020)\citenamefont {Dodero},
  \citenamefont {Vicini}, \citenamefont {Alloisio},\ and\ \citenamefont
  {Castellano}}]{Dodero_2020}%
  \BibitemOpen
  \bibfield  {author} {\bibinfo {author} {\bibfnamefont {A.}~\bibnamefont
  {Dodero}}, \bibinfo {author} {\bibfnamefont {S.}~\bibnamefont {Vicini}},
  \bibinfo {author} {\bibfnamefont {M.}~\bibnamefont {Alloisio}},\ and\
  \bibinfo {author} {\bibfnamefont {M.}~\bibnamefont {Castellano}},\ }\bibfield
   {title} {\bibinfo {title} {Rheological properties of sodium alginate
  solutions in the presence of added salt: an application of kulicke
  equation},\ }\href@noop {} {\bibfield  {journal} {\bibinfo  {journal}
  {Rheologica Acta}\ }\textbf {\bibinfo {volume} {59}},\ \bibinfo {pages} {365}
  (\bibinfo {year} {2020})}\BibitemShut {NoStop}%
\bibitem [{\citenamefont {Godefroid}\ \emph {et~al.}(2025)\citenamefont
  {Godefroid}, \citenamefont {Bouttes}, \citenamefont {Pereira},\ and\
  \citenamefont {Monteux}}]{Godefroid_2025}%
  \BibitemOpen
  \bibfield  {author} {\bibinfo {author} {\bibfnamefont {J.}~\bibnamefont
  {Godefroid}}, \bibinfo {author} {\bibfnamefont {D.}~\bibnamefont {Bouttes}},
  \bibinfo {author} {\bibfnamefont {A.}~\bibnamefont {Pereira}},\ and\ \bibinfo
  {author} {\bibfnamefont {C.}~\bibnamefont {Monteux}},\ }\bibfield  {title}
  {\bibinfo {title} {Gelation effects on the spreading of non-{N}ewtonian drops
  impacting a reactive liquid},\ }\href@noop {} {\bibfield  {journal} {\bibinfo
   {journal} {Soft Matter}\ }\textbf {\bibinfo {volume} {21}},\ \bibinfo
  {pages} {9162} (\bibinfo {year} {2025})}\BibitemShut {NoStop}%
\bibitem [{\citenamefont {Murphy}\ and\ \citenamefont
  {Atala}(2014)}]{Murphy_2014}%
  \BibitemOpen
  \bibfield  {author} {\bibinfo {author} {\bibfnamefont {V.}~\bibnamefont
  {Murphy}}\ and\ \bibinfo {author} {\bibfnamefont {A.}~\bibnamefont {Atala}},\
  }\bibfield  {title} {\bibinfo {title} {3d bioprinting of tissues and
  organs},\ }\href@noop {} {\bibfield  {journal} {\bibinfo  {journal} {Nature
  Biotechnology}\ }\textbf {\bibinfo {volume} {32}},\ \bibinfo {pages} {773}
  (\bibinfo {year} {2014})}\BibitemShut {NoStop}%
\bibitem [{\citenamefont {Yuk}(2018)}]{Bochenek_2018}%
  \BibitemOpen
  \bibfield  {author} {\bibinfo {author} {\bibfnamefont {H.}~\bibnamefont
  {Yuk}},\ }\bibfield  {title} {\bibinfo {title} {Alginate encapsulation as
  long-term immune protection of allogeneic pancreatic islet cells transplanted
  into the omental bursa of macaques},\ }\href@noop {} {\bibfield  {journal}
  {\bibinfo  {journal} {Nature Biomedical Engineering}\ }\textbf {\bibinfo
  {volume} {2}},\ \bibinfo {pages} {810–821} (\bibinfo {year}
  {2018})}\BibitemShut {NoStop}%
\bibitem [{\citenamefont {Yuk}(2022)}]{Yuk_2022}%
  \BibitemOpen
  \bibfield  {author} {\bibinfo {author} {\bibfnamefont {H.}~\bibnamefont
  {Yuk}},\ }\bibfield  {title} {\bibinfo {title} {Hydrogel interfaces for
  merging humans and machines},\ }\href@noop {} {\bibfield  {journal} {\bibinfo
   {journal} {Nature Reviews Materials}\ }\textbf {\bibinfo {volume} {7}},\
  \bibinfo {pages} {935} (\bibinfo {year} {2022})}\BibitemShut {NoStop}%
\bibitem [{\citenamefont {Ji}\ \emph {et~al.}(2022)\citenamefont {Ji},
  \citenamefont {Park}, \citenamefont {Oh}, \citenamefont {Nguyen},
  \citenamefont {Shin}, \citenamefont {Kim}, \citenamefont {Kim}, \citenamefont
  {Park},\ and\ \citenamefont {Kim}}]{Ji_2022}%
  \BibitemOpen
  \bibfield  {author} {\bibinfo {author} {\bibfnamefont {D.}~\bibnamefont
  {Ji}}, \bibinfo {author} {\bibfnamefont {J.~M.}\ \bibnamefont {Park}},
  \bibinfo {author} {\bibfnamefont {M.~S.}\ \bibnamefont {Oh}}, \bibinfo
  {author} {\bibfnamefont {T.~L.}\ \bibnamefont {Nguyen}}, \bibinfo {author}
  {\bibfnamefont {H.}~\bibnamefont {Shin}}, \bibinfo {author} {\bibfnamefont
  {J.~S.}\ \bibnamefont {Kim}}, \bibinfo {author} {\bibfnamefont
  {D.}~\bibnamefont {Kim}}, \bibinfo {author} {\bibfnamefont {H.~S.}\
  \bibnamefont {Park}},\ and\ \bibinfo {author} {\bibfnamefont
  {J.}~\bibnamefont {Kim}},\ }\bibfield  {title} {\bibinfo {title}
  {Superstrong, superstiff, and conductive alginate hydrogels},\ }\href@noop {}
  {\bibfield  {journal} {\bibinfo  {journal} {Nature Communcations}\ }\textbf
  {\bibinfo {volume} {13}},\ \bibinfo {pages} {1–10} (\bibinfo {year}
  {2022})}\BibitemShut {NoStop}%
\bibitem [{\citenamefont {Liang}\ and\ \citenamefont
  {Mackley}(1994)}]{Liang_1994}%
  \BibitemOpen
  \bibfield  {author} {\bibinfo {author} {\bibfnamefont {R.~F.}\ \bibnamefont
  {Liang}}\ and\ \bibinfo {author} {\bibfnamefont {M.~R.}\ \bibnamefont
  {Mackley}},\ }\bibfield  {title} {\bibinfo {title} {Rheological
  characterization of the time and strain dependence for polyisobutylene
  solutions},\ }\href@noop {} {\bibfield  {journal} {\bibinfo  {journal}
  {Journal of Non-Newtonian Fluid Mechanics}\ }\textbf {\bibinfo {volume}
  {52}},\ \bibinfo {pages} {387} (\bibinfo {year} {1994})}\BibitemShut
  {NoStop}%
\bibitem [{\citenamefont {Entov}\ and\ \citenamefont
  {Hinch}(1997)}]{Entov_1997}%
  \BibitemOpen
  \bibfield  {author} {\bibinfo {author} {\bibfnamefont {V.~M.}\ \bibnamefont
  {Entov}}\ and\ \bibinfo {author} {\bibfnamefont {E.~J.}\ \bibnamefont
  {Hinch}},\ }\bibfield  {title} {\bibinfo {title} {Effect of a spectrum of
  relaxation times on the capillary thinning of a filament of elastic liquid},\
  }\href@noop {} {\bibfield  {journal} {\bibinfo  {journal} {Journal of
  Non-Newtonian Fluid Mechanics}\ }\textbf {\bibinfo {volume} {72}},\ \bibinfo
  {pages} {31} (\bibinfo {year} {1997})}\BibitemShut {NoStop}%
\bibitem [{\citenamefont {McKinley}\ and\ \citenamefont
  {Tripathi}(2000)}]{McKinley_2000}%
  \BibitemOpen
  \bibfield  {author} {\bibinfo {author} {\bibfnamefont {G.~H.}\ \bibnamefont
  {McKinley}}\ and\ \bibinfo {author} {\bibfnamefont {A.}~\bibnamefont
  {Tripathi}},\ }\bibfield  {title} {\bibinfo {title} {How to extract the
  newtonian viscosity from capillary breakup measurements in a filament
  rheometer},\ }\href@noop {} {\bibfield  {journal} {\bibinfo  {journal}
  {Journal of Rheology}\ }\textbf {\bibinfo {volume} {44}},\ \bibinfo {pages}
  {653} (\bibinfo {year} {2000})}\BibitemShut {NoStop}%
\bibitem [{\citenamefont {Tuladhar}\ and\ \citenamefont
  {Mackley}(2008)}]{Tuladhar_2008}%
  \BibitemOpen
  \bibfield  {author} {\bibinfo {author} {\bibfnamefont {T.~R.}\ \bibnamefont
  {Tuladhar}}\ and\ \bibinfo {author} {\bibfnamefont {M.~R.}\ \bibnamefont
  {Mackley}},\ }\bibfield  {title} {\bibinfo {title} {Filament stretching
  rheometry and break-up behaviour of low viscosity polymer solutions and
  inkjet fluids},\ }\href@noop {} {\bibfield  {journal} {\bibinfo  {journal}
  {Journal of Non-Newtonian Fluid Mechanics}\ }\textbf {\bibinfo {volume}
  {148}},\ \bibinfo {pages} {97} (\bibinfo {year} {2008})}\BibitemShut
  {NoStop}%
\bibitem [{\citenamefont {Keshavarz}\ \emph {et~al.}(2015)\citenamefont
  {Keshavarz}, \citenamefont {Sharma}, \citenamefont {Houze}, \citenamefont
  {Koerner}, \citenamefont {Moore}, \citenamefont {Cotts}, \citenamefont
  {{T}hrelfall {H}olmes},\ and\ \citenamefont {McKinley}}]{Keshavarz_2015}%
  \BibitemOpen
  \bibfield  {author} {\bibinfo {author} {\bibfnamefont {B.}~\bibnamefont
  {Keshavarz}}, \bibinfo {author} {\bibfnamefont {V.}~\bibnamefont {Sharma}},
  \bibinfo {author} {\bibfnamefont {E.~C.}\ \bibnamefont {Houze}}, \bibinfo
  {author} {\bibfnamefont {M.~R.}\ \bibnamefont {Koerner}}, \bibinfo {author}
  {\bibfnamefont {J.~R.}\ \bibnamefont {Moore}}, \bibinfo {author}
  {\bibfnamefont {P.~M.}\ \bibnamefont {Cotts}}, \bibinfo {author}
  {\bibfnamefont {P.}~\bibnamefont {{T}hrelfall {H}olmes}},\ and\ \bibinfo
  {author} {\bibfnamefont {G.~H.}\ \bibnamefont {McKinley}},\ }\bibfield
  {title} {\bibinfo {title} {Studying the effects of elongational properties on
  atomization of weakly viscoelastic solutions using {R}ayleigh {O}hnesorge
  jetting extensional rheometry ({ROJER})},\ }\href@noop {} {\bibfield
  {journal} {\bibinfo  {journal} {Journal of Non-Newtonian Fluid Mechanics}\
  }\textbf {\bibinfo {volume} {222}},\ \bibinfo {pages} {171} (\bibinfo {year}
  {2015})}\BibitemShut {NoStop}%
\bibitem [{\citenamefont {Zinelis}\ \emph {et~al.}(2024)\citenamefont
  {Zinelis}, \citenamefont {Abadie}, \citenamefont {McKinley},\ and\
  \citenamefont {Matar}}]{Zinelis_2024}%
  \BibitemOpen
  \bibfield  {author} {\bibinfo {author} {\bibfnamefont {K.}~\bibnamefont
  {Zinelis}}, \bibinfo {author} {\bibfnamefont {T.}~\bibnamefont {Abadie}},
  \bibinfo {author} {\bibfnamefont {G.~H.}\ \bibnamefont {McKinley}},\ and\
  \bibinfo {author} {\bibfnamefont {O.~K.}\ \bibnamefont {Matar}},\ }\bibfield
  {title} {\bibinfo {title} {The fluid dynamics of a viscoelastic fluid
  dripping onto a substrate},\ }\href@noop {} {\bibfield  {journal} {\bibinfo
  {journal} {Soft Matter}\ }\textbf {\bibinfo {volume} {20}},\ \bibinfo {pages}
  {198} (\bibinfo {year} {2024})}\BibitemShut {NoStop}%
\bibitem [{\citenamefont {{El Khoury}}\ \emph {et~al.}(2026)\citenamefont {{El
  Khoury}}, \citenamefont {Isukwem}, \citenamefont {Hachem},\ and\
  \citenamefont {Pereira}}]{Khoury_2026}%
  \BibitemOpen
  \bibfield  {author} {\bibinfo {author} {\bibfnamefont {R.}~\bibnamefont {{El
  Khoury}}}, \bibinfo {author} {\bibfnamefont {K.}~\bibnamefont {Isukwem}},
  \bibinfo {author} {\bibfnamefont {E.}~\bibnamefont {Hachem}},\ and\ \bibinfo
  {author} {\bibfnamefont {A.}~\bibnamefont {Pereira}},\ }\bibfield  {title}
  {\bibinfo {title} {Dripping-onto-droplet capillary breakup},\ }\href@noop {}
  {\bibfield  {journal} {\bibinfo  {journal} {arXiv}\ ,\ \bibinfo {pages} {1}}
  (\bibinfo {year} {2026})}\BibitemShut {NoStop}%
\bibitem [{\citenamefont {Plateau}(1873)}]{Plateau_1873}%
  \BibitemOpen
  \bibfield  {author} {\bibinfo {author} {\bibfnamefont {J.}~\bibnamefont
  {Plateau}},\ }\bibfield  {title} {\bibinfo {title} {Statique exp\'erimentale
  et th\'eorique des liquides soumis aux seules forces mol\'eculaires},\
  }\href@noop {} {\bibfield  {journal} {\bibinfo  {journal} {Paris,
  Gauthier-Villars}\ } (\bibinfo {year} {1873})}\BibitemShut {NoStop}%
\bibitem [{\citenamefont {Rayleigh}(1878)}]{Rayleigh_1878}%
  \BibitemOpen
  \bibfield  {author} {\bibinfo {author} {\bibfnamefont {L.}~\bibnamefont
  {Rayleigh}},\ }\bibfield  {title} {\bibinfo {title} {The theory of sound -
  volume 2},\ }\href@noop {} {\bibfield  {journal} {\bibinfo  {journal}
  {London: Macmillan and Co}\ ,\ \bibinfo {pages} {337–398}} (\bibinfo {year}
  {1878})}\BibitemShut {NoStop}%
\bibitem [{\citenamefont {Rayleigh}(1880)}]{Rayleigh_1880}%
  \BibitemOpen
  \bibfield  {author} {\bibinfo {author} {\bibfnamefont {L.}~\bibnamefont
  {Rayleigh}},\ }\bibfield  {title} {\bibinfo {title} {On the stability, or
  instability, of certain fluid motions},\ }\href@noop {} {\bibfield  {journal}
  {\bibinfo  {journal} {Proceedings of the London Mathematical Society}\
  }\textbf {\bibinfo {volume} {s1-11}},\ \bibinfo {pages} {57} (\bibinfo {year}
  {1880})}\BibitemShut {NoStop}%
\bibitem [{\citenamefont {Rayleigh}(1892)}]{Rayleigh_1892}%
  \BibitemOpen
  \bibfield  {author} {\bibinfo {author} {\bibfnamefont {L.}~\bibnamefont
  {Rayleigh}},\ }\bibfield  {title} {\bibinfo {title} {On the instability of a
  cylinder of viscous liquid under capillary force},\ }\href@noop {} {\bibfield
   {journal} {\bibinfo  {journal} {The London, Edinburgh, and Dublin
  Philosophical Magazine and Journal of Science}\ }\textbf {\bibinfo {volume}
  {34}},\ \bibinfo {pages} {145} (\bibinfo {year} {1892})}\BibitemShut
  {NoStop}%
\bibitem [{\citenamefont {Gaudet}\ \emph {et~al.}(1996)\citenamefont {Gaudet},
  \citenamefont {McKinley},\ and\ \citenamefont {Stone}}]{Gaudet_1996}%
  \BibitemOpen
  \bibfield  {author} {\bibinfo {author} {\bibfnamefont {S.}~\bibnamefont
  {Gaudet}}, \bibinfo {author} {\bibfnamefont {G.~H.}\ \bibnamefont
  {McKinley}},\ and\ \bibinfo {author} {\bibfnamefont {H.~A.}\ \bibnamefont
  {Stone}},\ }\bibfield  {title} {\bibinfo {title} {Extensional deformation of
  newtonian liquid bridges},\ }\href@noop {} {\bibfield  {journal} {\bibinfo
  {journal} {Physics of Fluids}\ }\textbf {\bibinfo {volume} {8}},\ \bibinfo
  {pages} {2568} (\bibinfo {year} {1996})}\BibitemShut {NoStop}%
\bibitem [{\citenamefont {Anna}\ and\ \citenamefont
  {McKinley}(2001)}]{Anna_2001}%
  \BibitemOpen
  \bibfield  {author} {\bibinfo {author} {\bibfnamefont {S.~L.}\ \bibnamefont
  {Anna}}\ and\ \bibinfo {author} {\bibfnamefont {G.~H.}\ \bibnamefont
  {McKinley}},\ }\bibfield  {title} {\bibinfo {title} {Elasto-capillary
  thinning and breakup of model elastic liquids},\ }\href@noop {} {\bibfield
  {journal} {\bibinfo  {journal} {Journal of Rheology}\ }\textbf {\bibinfo
  {volume} {45}},\ \bibinfo {pages} {115} (\bibinfo {year} {2001})}\BibitemShut
  {NoStop}%
\bibitem [{\citenamefont {Rodd}\ \emph {et~al.}(2005)\citenamefont {Rodd},
  \citenamefont {Scott}, \citenamefont {Cooper-White},\ and\ \citenamefont
  {McKinley}}]{Rodd_2005}%
  \BibitemOpen
  \bibfield  {author} {\bibinfo {author} {\bibfnamefont {L.~E.}\ \bibnamefont
  {Rodd}}, \bibinfo {author} {\bibfnamefont {T.~P.}\ \bibnamefont {Scott}},
  \bibinfo {author} {\bibfnamefont {J.~J.}\ \bibnamefont {Cooper-White}},\ and\
  \bibinfo {author} {\bibfnamefont {G.~H.}\ \bibnamefont {McKinley}},\
  }\bibfield  {title} {\bibinfo {title} {Capillary break-up rheometry of
  low-viscosity elastic fluids},\ }\href@noop {} {\bibfield  {journal}
  {\bibinfo  {journal} {Applied Rheology}\ }\textbf {\bibinfo {volume} {15}},\
  \bibinfo {pages} {12} (\bibinfo {year} {2005})}\BibitemShut {NoStop}%
\bibitem [{\citenamefont {Clasen}\ \emph
  {et~al.}(2006{\natexlab{a}})\citenamefont {Clasen}, \citenamefont {Eggers},
  \citenamefont {Fontelos}, \citenamefont {Li},\ and\ \citenamefont
  {McKinley}}]{Clasen_2006b}%
  \BibitemOpen
  \bibfield  {author} {\bibinfo {author} {\bibfnamefont {C.}~\bibnamefont
  {Clasen}}, \bibinfo {author} {\bibfnamefont {J.}~\bibnamefont {Eggers}},
  \bibinfo {author} {\bibfnamefont {M.~A.}\ \bibnamefont {Fontelos}}, \bibinfo
  {author} {\bibfnamefont {J.}~\bibnamefont {Li}},\ and\ \bibinfo {author}
  {\bibfnamefont {G.~H.}\ \bibnamefont {McKinley}},\ }\bibfield  {title}
  {\bibinfo {title} {The beads-on-string structure of viscoelastic threads},\
  }\href@noop {} {\bibfield  {journal} {\bibinfo  {journal} {Journal of Fluid
  Mechanics}\ }\textbf {\bibinfo {volume} {556}},\ \bibinfo {pages} {283}
  (\bibinfo {year} {2006}{\natexlab{a}})}\BibitemShut {NoStop}%
\bibitem [{\citenamefont {Valette}\ \emph {et~al.}(2019)\citenamefont
  {Valette}, \citenamefont {Hachem}, \citenamefont {Khalloufi}, \citenamefont
  {Pereira}, \citenamefont {Mackley},\ and\ \citenamefont
  {Butler}}]{Valette_2019}%
  \BibitemOpen
  \bibfield  {author} {\bibinfo {author} {\bibfnamefont {R.}~\bibnamefont
  {Valette}}, \bibinfo {author} {\bibfnamefont {E.}~\bibnamefont {Hachem}},
  \bibinfo {author} {\bibfnamefont {M.}~\bibnamefont {Khalloufi}}, \bibinfo
  {author} {\bibfnamefont {A.}~\bibnamefont {Pereira}}, \bibinfo {author}
  {\bibfnamefont {M.}~\bibnamefont {Mackley}},\ and\ \bibinfo {author}
  {\bibfnamefont {S.}~\bibnamefont {Butler}},\ }\bibfield  {title} {\bibinfo
  {title} {The effect of viscosity, yield stress, and surface tension on the
  deformation and breakup profiles of fluid filaments stretched at very high
  velocities},\ }\href@noop {} {\bibfield  {journal} {\bibinfo  {journal}
  {Journal of Non-Newtonian Fluid Mechanics}\ }\textbf {\bibinfo {volume}
  {263}},\ \bibinfo {pages} {130} (\bibinfo {year} {2019})}\BibitemShut
  {NoStop}%
\bibitem [{\citenamefont {Joseph}\ and\ \citenamefont
  {Rothstein}(2025)}]{Joseph_2025}%
  \BibitemOpen
  \bibfield  {author} {\bibinfo {author} {\bibfnamefont {J.~B.}\ \bibnamefont
  {Joseph}}\ and\ \bibinfo {author} {\bibfnamefont {J.~P.}\ \bibnamefont
  {Rothstein}},\ }\bibfield  {title} {\bibinfo {title} {Recovery dynamics and
  polymer scission in capillary breakup extensional rheometry},\ }\href@noop {}
  {\bibfield  {journal} {\bibinfo  {journal} {Journal of Non-Newtonian Fluid
  Mechanics}\ }\textbf {\bibinfo {volume} {337}},\ \bibinfo {pages} {1}
  (\bibinfo {year} {2025})}\BibitemShut {NoStop}%
\bibitem [{\citenamefont {Dinic}\ \emph {et~al.}(2015)\citenamefont {Dinic},
  \citenamefont {Zhang}, \citenamefont {Jimenez},\ and\ \citenamefont
  {Sharma}}]{Dinic_2015}%
  \BibitemOpen
  \bibfield  {author} {\bibinfo {author} {\bibfnamefont {J.}~\bibnamefont
  {Dinic}}, \bibinfo {author} {\bibfnamefont {Y.}~\bibnamefont {Zhang}},
  \bibinfo {author} {\bibfnamefont {L.~N.}\ \bibnamefont {Jimenez}},\ and\
  \bibinfo {author} {\bibfnamefont {V.}~\bibnamefont {Sharma}},\ }\bibfield
  {title} {\bibinfo {title} {Extensional relaxation times of dilute, aqueous
  polymer solutions},\ }\href@noop {} {\bibfield  {journal} {\bibinfo
  {journal} {ACS Macro Letters}\ }\textbf {\bibinfo {volume} {4}},\ \bibinfo
  {pages} {804} (\bibinfo {year} {2015})}\BibitemShut {NoStop}%
\bibitem [{\citenamefont {Dinic}\ \emph {et~al.}(2017)\citenamefont {Dinic},
  \citenamefont {Jimenez},\ and\ \citenamefont {Sharma}}]{Dinic_2017}%
  \BibitemOpen
  \bibfield  {author} {\bibinfo {author} {\bibfnamefont {J.}~\bibnamefont
  {Dinic}}, \bibinfo {author} {\bibfnamefont {L.~N.}\ \bibnamefont {Jimenez}},\
  and\ \bibinfo {author} {\bibfnamefont {V.}~\bibnamefont {Sharma}},\
  }\bibfield  {title} {\bibinfo {title} {Pinch-off dynamics and
  dripping-onto-substrate (dos) rheometry of complex fluids},\ }\href@noop {}
  {\bibfield  {journal} {\bibinfo  {journal} {Lab on a Chip}\ }\textbf
  {\bibinfo {volume} {17}},\ \bibinfo {pages} {460} (\bibinfo {year}
  {2017})}\BibitemShut {NoStop}%
\bibitem [{\citenamefont {Rosello}\ \emph {et~al.}(2019)\citenamefont
  {Rosello}, \citenamefont {Sur}, \citenamefont {Barbet},\ and\ \citenamefont
  {Rothstein}}]{Rosello_2019}%
  \BibitemOpen
  \bibfield  {author} {\bibinfo {author} {\bibfnamefont {M.}~\bibnamefont
  {Rosello}}, \bibinfo {author} {\bibfnamefont {S.}~\bibnamefont {Sur}},
  \bibinfo {author} {\bibfnamefont {B.}~\bibnamefont {Barbet}},\ and\ \bibinfo
  {author} {\bibfnamefont {J.~P.}\ \bibnamefont {Rothstein}},\ }\bibfield
  {title} {\bibinfo {title} {Dripping-onto-substrate capillary breakup
  extensional rheometry of low-viscosity printing inks},\ }\href@noop {}
  {\bibfield  {journal} {\bibinfo  {journal} {Journal of Non-Newtonian Fluid
  Mechanics}\ }\textbf {\bibinfo {volume} {266}},\ \bibinfo {pages} {160}
  (\bibinfo {year} {2019})}\BibitemShut {NoStop}%
\bibitem [{\citenamefont {Bhattacharjee}\ \emph {et~al.}(2011)\citenamefont
  {Bhattacharjee}, \citenamefont {McDonnell}, \citenamefont {Prabhakar},
  \citenamefont {Yeo},\ and\ \citenamefont {Friend}}]{Bhattacharjee_2011}%
  \BibitemOpen
  \bibfield  {author} {\bibinfo {author} {\bibfnamefont {P.~K.}\ \bibnamefont
  {Bhattacharjee}}, \bibinfo {author} {\bibfnamefont {A.~G.}\ \bibnamefont
  {McDonnell}}, \bibinfo {author} {\bibfnamefont {R.}~\bibnamefont
  {Prabhakar}}, \bibinfo {author} {\bibfnamefont {L.~Y.}\ \bibnamefont {Yeo}},\
  and\ \bibinfo {author} {\bibfnamefont {J.}~\bibnamefont {Friend}},\
  }\bibfield  {title} {\bibinfo {title} {Extensional flow of low-viscosity
  fluids in capillary bridges formed by pulsed surface acoustic wave jetting},\
  }\href@noop {} {\bibfield  {journal} {\bibinfo  {journal} {New Journal of
  Physics}\ }\textbf {\bibinfo {volume} {13}},\ \bibinfo {pages} {1} (\bibinfo
  {year} {2011})}\BibitemShut {NoStop}%
\bibitem [{\citenamefont {McDonnell}\ \emph {et~al.}(2015)\citenamefont
  {McDonnell}, \citenamefont {Gopesh}, \citenamefont {Lo}, \citenamefont
  {O'Bryan}, \citenamefont {Yeo}, \citenamefont {Friend},\ and\ \citenamefont
  {Prabhakar}}]{McDonnell_2015}%
  \BibitemOpen
  \bibfield  {author} {\bibinfo {author} {\bibfnamefont {A.~G.}\ \bibnamefont
  {McDonnell}}, \bibinfo {author} {\bibfnamefont {T.~C.}\ \bibnamefont
  {Gopesh}}, \bibinfo {author} {\bibfnamefont {J.}~\bibnamefont {Lo}}, \bibinfo
  {author} {\bibfnamefont {M.}~\bibnamefont {O'Bryan}}, \bibinfo {author}
  {\bibfnamefont {L.}~\bibnamefont {Yeo}}, \bibinfo {author} {\bibfnamefont
  {J.~R.}\ \bibnamefont {Friend}},\ and\ \bibinfo {author} {\bibfnamefont
  {R.}~\bibnamefont {Prabhakar}},\ }\bibfield  {title} {\bibinfo {title}
  {Motility induced changes in viscosity of suspensions of swimming microbes in
  extensional flows},\ }\href@noop {} {\bibfield  {journal} {\bibinfo
  {journal} {Soft Matter}\ }\textbf {\bibinfo {volume} {11}},\ \bibinfo {pages}
  {4658} (\bibinfo {year} {2015})}\BibitemShut {NoStop}%
\bibitem [{\citenamefont {Oldroyd}(1950)}]{Oldroyd_1950}%
  \BibitemOpen
  \bibfield  {author} {\bibinfo {author} {\bibfnamefont {J.~G.}\ \bibnamefont
  {Oldroyd}},\ }\bibfield  {title} {\bibinfo {title} {On the formulation of
  rheological equations of state},\ }\href@noop {} {\bibfield  {journal}
  {\bibinfo  {journal} {Proceedings of the Royal Society A}\ }\textbf {\bibinfo
  {volume} {200}},\ \bibinfo {pages} {523} (\bibinfo {year}
  {1950})}\BibitemShut {NoStop}%
\bibitem [{\citenamefont {Bird}\ \emph {et~al.}(1987)\citenamefont {Bird},
  \citenamefont {Armstrong},\ and\ \citenamefont {O.}}]{Bird_1987}%
  \BibitemOpen
  \bibfield  {author} {\bibinfo {author} {\bibfnamefont {R.~B.}\ \bibnamefont
  {Bird}}, \bibinfo {author} {\bibfnamefont {R.~C.}\ \bibnamefont
  {Armstrong}},\ and\ \bibinfo {author} {\bibfnamefont {H.}~\bibnamefont
  {O.}},\ }\bibfield  {title} {\bibinfo {title} {Dynamics of polymeric
  liquids},\ }\href@noop {} {\bibfield  {journal} {\bibinfo  {journal}
  {Wiley-Interscience, New York}\ }\textbf {\bibinfo {volume} {2nd edition}},\
  \bibinfo {pages} {172} (\bibinfo {year} {1987})}\BibitemShut {NoStop}%
\bibitem [{\citenamefont {Hinch}\ and\ \citenamefont
  {Harlen}(2021)}]{Hinch_2021}%
  \BibitemOpen
  \bibfield  {author} {\bibinfo {author} {\bibfnamefont {J.}~\bibnamefont
  {Hinch}}\ and\ \bibinfo {author} {\bibfnamefont {O.}~\bibnamefont {Harlen}},\
  }\bibfield  {title} {\bibinfo {title} {Oldroyd {B}, and not {A}?},\
  }\href@noop {} {\bibfield  {journal} {\bibinfo  {journal} {Journal of
  Non-Newtonian Fluid Mechanics}\ }\textbf {\bibinfo {volume} {298}},\ \bibinfo
  {pages} {1} (\bibinfo {year} {2021})}\BibitemShut {NoStop}%
\bibitem [{\citenamefont {Bazilevskii}\ \emph {et~al.}(1997)\citenamefont
  {Bazilevskii}, \citenamefont {Entov}, \citenamefont {Lerner},\ and\
  \citenamefont {Rozhkov}}]{Bazilevskii_1997}%
  \BibitemOpen
  \bibfield  {author} {\bibinfo {author} {\bibfnamefont {A.~V.}\ \bibnamefont
  {Bazilevskii}}, \bibinfo {author} {\bibfnamefont {V.~M.}\ \bibnamefont
  {Entov}}, \bibinfo {author} {\bibfnamefont {M.~M.}\ \bibnamefont {Lerner}},\
  and\ \bibinfo {author} {\bibfnamefont {A.~N.}\ \bibnamefont {Rozhkov}},\
  }\bibfield  {title} {\bibinfo {title} {Failure of polymer solution
  filaments},\ }\href@noop {} {\bibfield  {journal} {\bibinfo  {journal}
  {Polymer Science - Series A}\ }\textbf {\bibinfo {volume} {39}},\ \bibinfo
  {pages} {316} (\bibinfo {year} {1997})}\BibitemShut {NoStop}%
\bibitem [{\citenamefont {Clasen}\ \emph
  {et~al.}(2006{\natexlab{b}})\citenamefont {Clasen}, \citenamefont {Eggers},
  \citenamefont {Fontelos}, \citenamefont {Li},\ and\ \citenamefont
  {McKinley}}]{Clasen_2006a}%
  \BibitemOpen
  \bibfield  {author} {\bibinfo {author} {\bibfnamefont {C.}~\bibnamefont
  {Clasen}}, \bibinfo {author} {\bibfnamefont {J.}~\bibnamefont {Eggers}},
  \bibinfo {author} {\bibfnamefont {M.~A.}\ \bibnamefont {Fontelos}}, \bibinfo
  {author} {\bibfnamefont {J.}~\bibnamefont {Li}},\ and\ \bibinfo {author}
  {\bibfnamefont {G.~H.}\ \bibnamefont {McKinley}},\ }\bibfield  {title}
  {\bibinfo {title} {The beads-on-string structure of viscoelastic threads},\
  }\href@noop {} {\bibfield  {journal} {\bibinfo  {journal} {Journal of Fluid
  Mechanics}\ }\textbf {\bibinfo {volume} {556}},\ \bibinfo {pages} {283}
  (\bibinfo {year} {2006}{\natexlab{b}})}\BibitemShut {NoStop}%
\bibitem [{\citenamefont {Pereira}\ \emph {et~al.}(2013)\citenamefont
  {Pereira}, \citenamefont {Andrade},\ and\ \citenamefont
  {Soares}}]{Pereira_2013}%
  \BibitemOpen
  \bibfield  {author} {\bibinfo {author} {\bibfnamefont {A.}~\bibnamefont
  {Pereira}}, \bibinfo {author} {\bibfnamefont {R.~M.}\ \bibnamefont
  {Andrade}},\ and\ \bibinfo {author} {\bibfnamefont {E.~J.}\ \bibnamefont
  {Soares}},\ }\bibfield  {title} {\bibinfo {title} {Drag reduction induced by
  flexible and rigid molecules in a turbulent flow into a rotating cylindrical
  double gap device: Comparison between poly(ethylene oxide), polyacrylamide,
  and xanthan gum},\ }\href@noop {} {\bibfield  {journal} {\bibinfo  {journal}
  {Journal of Non-Newtonian Fluid Mechanics}\ }\textbf {\bibinfo {volume}
  {202}},\ \bibinfo {pages} {72} (\bibinfo {year} {2013})}\BibitemShut
  {NoStop}%
\bibitem [{\citenamefont {Eggers}(1993)}]{Eggers_1993}%
  \BibitemOpen
  \bibfield  {author} {\bibinfo {author} {\bibfnamefont {J.}~\bibnamefont
  {Eggers}},\ }\bibfield  {title} {\bibinfo {title} {Universal pinching of {3D}
  axisymmetric free-surface flow},\ }\href@noop {} {\bibfield  {journal}
  {\bibinfo  {journal} {Physical Review Letters}\ }\textbf {\bibinfo {volume}
  {71}},\ \bibinfo {pages} {3458} (\bibinfo {year} {1993})}\BibitemShut
  {NoStop}%
\bibitem [{\citenamefont {Eggers}\ and\ \citenamefont
  {Villermaux}(2008)}]{Eggers_2008}%
  \BibitemOpen
  \bibfield  {author} {\bibinfo {author} {\bibfnamefont {J.}~\bibnamefont
  {Eggers}}\ and\ \bibinfo {author} {\bibfnamefont {E.}~\bibnamefont
  {Villermaux}},\ }\bibfield  {title} {\bibinfo {title} {Physics of liquid
  jets},\ }\href@noop {} {\bibfield  {journal} {\bibinfo  {journal} {Reports on
  Progress in Physics}\ }\textbf {\bibinfo {volume} {71}},\ \bibinfo {pages}
  {1} (\bibinfo {year} {2008})}\BibitemShut {NoStop}%
\bibitem [{\citenamefont {Castrej\'on-Pita}\ \emph {et~al.}(2012)\citenamefont
  {Castrej\'on-Pita}, \citenamefont {Castrej\'on-Pita},\ and\ \citenamefont
  {Hutchings}}]{Pita_2012}%
  \BibitemOpen
  \bibfield  {author} {\bibinfo {author} {\bibfnamefont {A.~A.}\ \bibnamefont
  {Castrej\'on-Pita}}, \bibinfo {author} {\bibfnamefont {J.~R.}\ \bibnamefont
  {Castrej\'on-Pita}},\ and\ \bibinfo {author} {\bibfnamefont {I.}~\bibnamefont
  {Hutchings}},\ }\bibfield  {title} {\bibinfo {title} {The breakup of liquid
  filament},\ }\href@noop {} {\bibfield  {journal} {\bibinfo  {journal}
  {Physical Review Letters}\ }\textbf {\bibinfo {volume} {108}},\ \bibinfo
  {pages} {1} (\bibinfo {year} {2012})}\BibitemShut {NoStop}%
\bibitem [{\citenamefont {Deblais}\ \emph {et~al.}(2018)\citenamefont
  {Deblais}, \citenamefont {Herrada}, \citenamefont {Hauner}, \citenamefont
  {Velikov}, \citenamefont {van Roon}, \citenamefont {Kellay}, \citenamefont
  {Eggers},\ and\ \citenamefont {Bonn}}]{Deblais_2018}%
  \BibitemOpen
  \bibfield  {author} {\bibinfo {author} {\bibfnamefont {A.}~\bibnamefont
  {Deblais}}, \bibinfo {author} {\bibfnamefont {M.~A.}\ \bibnamefont
  {Herrada}}, \bibinfo {author} {\bibfnamefont {I.}~\bibnamefont {Hauner}},
  \bibinfo {author} {\bibfnamefont {K.~P.}\ \bibnamefont {Velikov}}, \bibinfo
  {author} {\bibfnamefont {T.}~\bibnamefont {van Roon}}, \bibinfo {author}
  {\bibfnamefont {H.}~\bibnamefont {Kellay}}, \bibinfo {author} {\bibfnamefont
  {J.}~\bibnamefont {Eggers}},\ and\ \bibinfo {author} {\bibfnamefont
  {D.}~\bibnamefont {Bonn}},\ }\bibfield  {title} {\bibinfo {title} {Viscous
  effects on inertial drop formation},\ }\href@noop {} {\bibfield  {journal}
  {\bibinfo  {journal} {Physical Review Letters}\ }\textbf {\bibinfo {volume}
  {121}},\ \bibinfo {pages} {254501} (\bibinfo {year} {2018})}\BibitemShut
  {NoStop}%
\bibitem [{\citenamefont {Kooij}\ \emph {et~al.}(2025)\citenamefont {Kooij},
  \citenamefont {Jordan}, \citenamefont {van Rijn}, \citenamefont {Ribe},\ and\
  \citenamefont {Bonn}}]{Kooij_2025}%
  \BibitemOpen
  \bibfield  {author} {\bibinfo {author} {\bibfnamefont {S.}~\bibnamefont
  {Kooij}}, \bibinfo {author} {\bibfnamefont {D.~T.~A.}\ \bibnamefont
  {Jordan}}, \bibinfo {author} {\bibfnamefont {C.~J.~M.}\ \bibnamefont {van
  Rijn}}, \bibinfo {author} {\bibfnamefont {N.~M.}\ \bibnamefont {Ribe}},\ and\
  \bibinfo {author} {\bibfnamefont {D.}~\bibnamefont {Bonn}},\ }\bibfield
  {title} {\bibinfo {title} {What determines the breakup length of a jet?},\
  }\href@noop {} {\bibfield  {journal} {\bibinfo  {journal} {Physical Review
  Letters}\ }\textbf {\bibinfo {volume} {135}},\ \bibinfo {pages} {214001}
  (\bibinfo {year} {2025})}\BibitemShut {NoStop}%
\bibitem [{\citenamefont {Kulicke}\ \emph {et~al.}(1982)\citenamefont
  {Kulicke}, \citenamefont {Kniewske},\ and\ \citenamefont
  {Klein}}]{Kulicke_1982}%
  \BibitemOpen
  \bibfield  {author} {\bibinfo {author} {\bibfnamefont {W.-M.}\ \bibnamefont
  {Kulicke}}, \bibinfo {author} {\bibfnamefont {R.}~\bibnamefont {Kniewske}},\
  and\ \bibinfo {author} {\bibfnamefont {J.}~\bibnamefont {Klein}},\ }\bibfield
   {title} {\bibinfo {title} {Preparation, characterization, solution
  properties and rheological behaviour of polyacrylamide},\ }\href@noop {}
  {\bibfield  {journal} {\bibinfo  {journal} {Progress in Polymer Science}\
  }\textbf {\bibinfo {volume} {8}},\ \bibinfo {pages} {373} (\bibinfo {year}
  {1982})}\BibitemShut {NoStop}%
\bibitem [{\citenamefont {Kulicke}\ and\ \citenamefont
  {Kniewske}(1984)}]{Kulicke_1984}%
  \BibitemOpen
  \bibfield  {author} {\bibinfo {author} {\bibfnamefont {W.-M.}\ \bibnamefont
  {Kulicke}}\ and\ \bibinfo {author} {\bibfnamefont {R.}~\bibnamefont
  {Kniewske}},\ }\bibfield  {title} {\bibinfo {title} {The shear viscosity
  dependence on concentration, molecular weight, and shear rate of polystyrene
  solutions},\ }\href@noop {} {\bibfield  {journal} {\bibinfo  {journal}
  {Rheologica Acta}\ }\textbf {\bibinfo {volume} {23}},\ \bibinfo {pages} {75}
  (\bibinfo {year} {1984})}\BibitemShut {NoStop}%
\bibitem [{\citenamefont {Xu}\ \emph {et~al.}(2014)\citenamefont {Xu},
  \citenamefont {Chen},\ and\ \citenamefont {An}}]{Xu_2014}%
  \BibitemOpen
  \bibfield  {author} {\bibinfo {author} {\bibfnamefont {X.}~\bibnamefont
  {Xu}}, \bibinfo {author} {\bibfnamefont {J.}~\bibnamefont {Chen}},\ and\
  \bibinfo {author} {\bibfnamefont {L.}~\bibnamefont {An}},\ }\bibfield
  {title} {\bibinfo {title} {Shear thinning behavior of linear polymer melts
  under shear flow via nonequilibrium molecular dynamics},\ }\href@noop {}
  {\bibfield  {journal} {\bibinfo  {journal} {The Journal of Chemical Physics}\
  }\textbf {\bibinfo {volume} {140}},\ \bibinfo {pages} {174902} (\bibinfo
  {year} {2014})}\BibitemShut {NoStop}%
\bibitem [{\citenamefont {Torres}\ \emph {et~al.}(2014)\citenamefont {Torres},
  \citenamefont {Hallmark},\ and\ \citenamefont {Wilson}}]{Torres_2014}%
  \BibitemOpen
  \bibfield  {author} {\bibinfo {author} {\bibfnamefont {M.~D.}\ \bibnamefont
  {Torres}}, \bibinfo {author} {\bibfnamefont {B.}~\bibnamefont {Hallmark}},\
  and\ \bibinfo {author} {\bibfnamefont {D.~I.}\ \bibnamefont {Wilson}},\
  }\bibfield  {title} {\bibinfo {title} {Effect of concentration on shear and
  extensional rheology of guar gum solutions},\ }\href@noop {} {\bibfield
  {journal} {\bibinfo  {journal} {Food Hydrocolloids}\ }\textbf {\bibinfo
  {volume} {40}},\ \bibinfo {pages} {85} (\bibinfo {year} {2014})}\BibitemShut
  {NoStop}%
\bibitem [{\citenamefont {Costanzo}\ \emph {et~al.}(2016)\citenamefont
  {Costanzo}, \citenamefont {Huang}, \citenamefont {Ianniruberto},
  \citenamefont {Marrucci}, \citenamefont {Hassager},\ and\ \citenamefont
  {Vlassopoulos}}]{Costanzo_2016}%
  \BibitemOpen
  \bibfield  {author} {\bibinfo {author} {\bibfnamefont {S.}~\bibnamefont
  {Costanzo}}, \bibinfo {author} {\bibfnamefont {Q.}~\bibnamefont {Huang}},
  \bibinfo {author} {\bibfnamefont {G.}~\bibnamefont {Ianniruberto}}, \bibinfo
  {author} {\bibfnamefont {G.}~\bibnamefont {Marrucci}}, \bibinfo {author}
  {\bibfnamefont {O.}~\bibnamefont {Hassager}},\ and\ \bibinfo {author}
  {\bibfnamefont {D.}~\bibnamefont {Vlassopoulos}},\ }\bibfield  {title}
  {\bibinfo {title} {Shear and extensional rheology of polystyrene melts and
  solutions with the same number of entanglements},\ }\href@noop {} {\bibfield
  {journal} {\bibinfo  {journal} {Macromolecules}\ }\textbf {\bibinfo {volume}
  {49}},\ \bibinfo {pages} {A} (\bibinfo {year} {2016})}\BibitemShut {NoStop}%
\bibitem [{\citenamefont {Bertasa}\ \emph {et~al.}(2020)\citenamefont
  {Bertasa}, \citenamefont {Dodero}, \citenamefont {Alloisio}, \citenamefont
  {Vicini}, \citenamefont {Riedo}, \citenamefont {Sansonetti}, \citenamefont
  {Scalarone},\ and\ \citenamefont {Castellano}}]{Bertasa_2020}%
  \BibitemOpen
  \bibfield  {author} {\bibinfo {author} {\bibfnamefont {M.}~\bibnamefont
  {Bertasa}}, \bibinfo {author} {\bibfnamefont {A.}~\bibnamefont {Dodero}},
  \bibinfo {author} {\bibfnamefont {M.}~\bibnamefont {Alloisio}}, \bibinfo
  {author} {\bibfnamefont {S.}~\bibnamefont {Vicini}}, \bibinfo {author}
  {\bibfnamefont {C.}~\bibnamefont {Riedo}}, \bibinfo {author} {\bibfnamefont
  {A.}~\bibnamefont {Sansonetti}}, \bibinfo {author} {\bibfnamefont
  {D.}~\bibnamefont {Scalarone}},\ and\ \bibinfo {author} {\bibfnamefont
  {M.}~\bibnamefont {Castellano}},\ }\bibfield  {title} {\bibinfo {title} {Agar
  gel strength: A correlation study between chemical composition and
  rheological properties},\ }\href@noop {} {\bibfield  {journal} {\bibinfo
  {journal} {European Polymer Journal}\ }\textbf {\bibinfo {volume} {123}},\
  \bibinfo {pages} {109442} (\bibinfo {year} {2020})}\BibitemShut {NoStop}%
\bibitem [{\citenamefont {Carreau}(1972)}]{Carreau_1972}%
  \BibitemOpen
  \bibfield  {author} {\bibinfo {author} {\bibfnamefont {P.~J.}\ \bibnamefont
  {Carreau}},\ }\bibfield  {title} {\bibinfo {title} {Rheological equations
  from molecular network theories},\ }\href@noop {} {\bibfield  {journal}
  {\bibinfo  {journal} {Transactions of the Society of Rheology}\ }\textbf
  {\bibinfo {volume} {16}},\ \bibinfo {pages} {99} (\bibinfo {year}
  {1972})}\BibitemShut {NoStop}%
\bibitem [{\citenamefont {Yasuda}(1979)}]{Yasuda_1979}%
  \BibitemOpen
  \bibfield  {author} {\bibinfo {author} {\bibfnamefont {K.}~\bibnamefont
  {Yasuda}},\ }\href@noop {} {\emph {\bibinfo {title} {Investigation of the
  analogies between viscometric and linear viscoelastic properties of
  polystyrene fluids}}}\ (\bibinfo  {publisher} {Ph.D. Thesis - Massachusetts
  Institute of Technology},\ \bibinfo {year} {1979})\BibitemShut {NoStop}%
\bibitem [{\citenamefont {Dobrynin}\ \emph {et~al.}(1995)\citenamefont
  {Dobrynin}, \citenamefont {Colby},\ and\ \citenamefont
  {Rubinstein}}]{Dobrynin_1995}%
  \BibitemOpen
  \bibfield  {author} {\bibinfo {author} {\bibfnamefont {A.~V.}\ \bibnamefont
  {Dobrynin}}, \bibinfo {author} {\bibfnamefont {R.~H.}\ \bibnamefont
  {Colby}},\ and\ \bibinfo {author} {\bibfnamefont {M.}~\bibnamefont
  {Rubinstein}},\ }\bibfield  {title} {\bibinfo {title} {Scaling theory of
  polyelectrolyte solutions},\ }\href@noop {} {\bibfield  {journal} {\bibinfo
  {journal} {Macromolecules}\ }\textbf {\bibinfo {volume} {28}},\ \bibinfo
  {pages} {1859} (\bibinfo {year} {1995})}\BibitemShut {NoStop}%
\bibitem [{\citenamefont {Kulicke}\ and\ \citenamefont
  {Clasen}(2004)}]{Kulicke_2004}%
  \BibitemOpen
  \bibfield  {author} {\bibinfo {author} {\bibfnamefont {W.-M.}\ \bibnamefont
  {Kulicke}}\ and\ \bibinfo {author} {\bibfnamefont {C.}~\bibnamefont
  {Clasen}},\ }\href@noop {} {\emph {\bibinfo {title} {Viscosimetry of polymers
  and polyelectrolytes}}}\ (\bibinfo  {publisher} {Springer Laboratory,
  Springer},\ \bibinfo {year} {2004})\BibitemShut {NoStop}%
\bibitem [{\citenamefont {Colby}(2010)}]{Colby_2010}%
  \BibitemOpen
  \bibfield  {author} {\bibinfo {author} {\bibfnamefont {R.~H.}\ \bibnamefont
  {Colby}},\ }\bibfield  {title} {\bibinfo {title} {Structure and linear
  viscoelasticity of flexible polymer solutions: comparison of polyelectrolyte
  and neutral polymer solutions},\ }\href@noop {} {\bibfield  {journal}
  {\bibinfo  {journal} {Rheologica Acta}\ }\textbf {\bibinfo {volume} {49}},\
  \bibinfo {pages} {425} (\bibinfo {year} {2010})}\BibitemShut {NoStop}%
\bibitem [{\citenamefont {Stauffer}(1965)}]{Stauffer_1965}%
  \BibitemOpen
  \bibfield  {author} {\bibinfo {author} {\bibfnamefont {C.~E.}\ \bibnamefont
  {Stauffer}},\ }\bibfield  {title} {\bibinfo {title} {The measurement of
  surface tension by the pendant drop technique},\ }\href@noop {} {\bibfield
  {journal} {\bibinfo  {journal} {The Journal of Physical Chemistry}\ }\textbf
  {\bibinfo {volume} {69}},\ \bibinfo {pages} {1933–1938} (\bibinfo {year}
  {1965})}\BibitemShut {NoStop}%
\bibitem [{\citenamefont {Adamson}\ and\ \citenamefont
  {Gast}(1997)}]{Adamson_1997}%
  \BibitemOpen
  \bibfield  {author} {\bibinfo {author} {\bibfnamefont {A.~W.}\ \bibnamefont
  {Adamson}}\ and\ \bibinfo {author} {\bibfnamefont {A.~P.}\ \bibnamefont
  {Gast}},\ }\href@noop {} {\emph {\bibinfo {title} {Physical chemistry of
  surfaces}}}\ (\bibinfo  {publisher} {John Wiley \& Sons},\ \bibinfo {year}
  {1997})\BibitemShut {NoStop}%
\bibitem [{\citenamefont {Ebnesajjad}(2011)}]{Ebnesajjad_2011}%
  \BibitemOpen
  \bibfield  {author} {\bibinfo {author} {\bibfnamefont {S.}~\bibnamefont
  {Ebnesajjad}},\ }\href@noop {} {\emph {\bibinfo {title} {Handbook of
  adhesives and surface preparation}}}\ (\bibinfo  {publisher} {Elsevier},\
  \bibinfo {year} {2011})\BibitemShut {NoStop}%
\bibitem [{\citenamefont {Berry}\ \emph {et~al.}(2015)\citenamefont {Berry},
  \citenamefont {Neeson}, \citenamefont {Dagastine}, \citenamefont {Chan},\
  and\ \citenamefont {Tabor}}]{Berry_2015}%
  \BibitemOpen
  \bibfield  {author} {\bibinfo {author} {\bibfnamefont {J.~D.}\ \bibnamefont
  {Berry}}, \bibinfo {author} {\bibfnamefont {M.~J.}\ \bibnamefont {Neeson}},
  \bibinfo {author} {\bibfnamefont {R.~R.}\ \bibnamefont {Dagastine}}, \bibinfo
  {author} {\bibfnamefont {D.~Y.}\ \bibnamefont {Chan}},\ and\ \bibinfo
  {author} {\bibfnamefont {R.~F.}\ \bibnamefont {Tabor}},\ }\bibfield  {title}
  {\bibinfo {title} {Measurement of surface and interfacial tension using
  pendant drop tensiometry},\ }\href@noop {} {\bibfield  {journal} {\bibinfo
  {journal} {Journal of Colloid and Interface Science}\ }\textbf {\bibinfo
  {volume} {16}},\ \bibinfo {pages} {226} (\bibinfo {year} {2015})}\BibitemShut
  {NoStop}%
\bibitem [{\citenamefont {Peregrine}\ \emph {et~al.}(1990)\citenamefont
  {Peregrine}, \citenamefont {Shoker},\ and\ \citenamefont
  {Symon}}]{Peregrine_1990}%
  \BibitemOpen
  \bibfield  {author} {\bibinfo {author} {\bibfnamefont {D.~H.}\ \bibnamefont
  {Peregrine}}, \bibinfo {author} {\bibfnamefont {G.}~\bibnamefont {Shoker}},\
  and\ \bibinfo {author} {\bibfnamefont {A.}~\bibnamefont {Symon}},\ }\bibfield
   {title} {\bibinfo {title} {The bifurcation of liquid bridges},\ }\href@noop
  {} {\bibfield  {journal} {\bibinfo  {journal} {Journal of Fluid Mechanics}\
  }\textbf {\bibinfo {volume} {212}},\ \bibinfo {pages} {25} (\bibinfo {year}
  {1990})}\BibitemShut {NoStop}%
\bibitem [{\citenamefont {Eggers}(2005)}]{Eggers_2005}%
  \BibitemOpen
  \bibfield  {author} {\bibinfo {author} {\bibfnamefont {J.}~\bibnamefont
  {Eggers}},\ }\bibfield  {title} {\bibinfo {title} {Drop formation - an
  overview},\ }\href@noop {} {\bibfield  {journal} {\bibinfo  {journal} {ZAMM -
  Journal of Applied Mathematics and Mechanics}\ }\textbf {\bibinfo {volume}
  {6}},\ \bibinfo {pages} {400–410} (\bibinfo {year} {2005})}\BibitemShut
  {NoStop}%
\bibitem [{\citenamefont {Javadi}\ \emph {et~al.}(2013)\citenamefont {Javadi},
  \citenamefont {Eggers}, \citenamefont {Bonn}, \citenamefont {Habibi},\ and\
  \citenamefont {Ribe}}]{Javadi_2013}%
  \BibitemOpen
  \bibfield  {author} {\bibinfo {author} {\bibfnamefont {A.}~\bibnamefont
  {Javadi}}, \bibinfo {author} {\bibfnamefont {J.}~\bibnamefont {Eggers}},
  \bibinfo {author} {\bibfnamefont {D.}~\bibnamefont {Bonn}}, \bibinfo {author}
  {\bibfnamefont {M.}~\bibnamefont {Habibi}},\ and\ \bibinfo {author}
  {\bibfnamefont {N.~M.}\ \bibnamefont {Ribe}},\ }\bibfield  {title} {\bibinfo
  {title} {Delayed capillary breakup of falling viscous jets},\ }\href@noop {}
  {\bibfield  {journal} {\bibinfo  {journal} {Physical Review Letters}\
  }\textbf {\bibinfo {volume} {110}},\ \bibinfo {pages} {144501} (\bibinfo
  {year} {2013})}\BibitemShut {NoStop}%
\bibitem [{\citenamefont {Dinic}\ and\ \citenamefont
  {Sharma}(2019)}]{Dinic_2019}%
  \BibitemOpen
  \bibfield  {author} {\bibinfo {author} {\bibfnamefont {J.}~\bibnamefont
  {Dinic}}\ and\ \bibinfo {author} {\bibfnamefont {V.}~\bibnamefont {Sharma}},\
  }\bibfield  {title} {\bibinfo {title} {Computational analysis of self-similar
  capillary-driven thinning and pinch-off dynamics during dripping using the
  volume-of-fluid method},\ }\href@noop {} {\bibfield  {journal} {\bibinfo
  {journal} {Physics of Fluids}\ }\textbf {\bibinfo {volume} {31}},\ \bibinfo
  {pages} {021211} (\bibinfo {year} {2019})}\BibitemShut {NoStop}%
\bibitem [{\citenamefont {Bazazi}\ and\ \citenamefont
  {Stone}(2025)}]{Bazazi_2025}%
  \BibitemOpen
  \bibfield  {author} {\bibinfo {author} {\bibfnamefont {P.}~\bibnamefont
  {Bazazi}}\ and\ \bibinfo {author} {\bibfnamefont {H.~A.}\ \bibnamefont
  {Stone}},\ }\bibfield  {title} {\bibinfo {title} {Pinch-off dynamics of
  emulsion filaments before and after polymerization of the internal phase},\
  }\href@noop {} {\bibfield  {journal} {\bibinfo  {journal} {Soft Matter}\
  }\textbf {\bibinfo {volume} {21}},\ \bibinfo {pages} {1296} (\bibinfo {year}
  {2025})}\BibitemShut {NoStop}%
\end{thebibliography}%
%%%%%%%%%%%%%%%%%%%%%%%%%%%%%%%%%%%%%%%%%%%%%%%%%%%%%%%%%%%%%%%%%%%%%%%%%%%%%%
%%%%%%%%%%%%%%%%%%%%%%%%%%%%%%%%%%%%%%%%%%%%%%%%%%%%%%%%%%%%%%%%%%%%%%%%%%%%%%
\end{document}